\documentclass[
    reprint,
    superscriptaddress,
    amsmath,amssymb,
    aps,
    prx,
    longbibliography,
]{revtex4-1}
\usepackage{braket}
\usepackage{graphicx}
\usepackage{amsmath}
\usepackage{amssymb}
\usepackage{algorithm}
\usepackage{algpseudocode}
\usepackage{bbm}
\usepackage[colorlinks=true,urlcolor=blue,linkcolor=blue,citecolor=blue]{hyperref}
\usepackage[margin=0.75in]{geometry}

\newcommand{\gs}{\mbox{\tiny gs}}
\newcommand{\x}{\mathbf{x}}
\newcommand{\Tr}{\operatorname{Tr}}
\newcommand{\unity}{\mathbbm{1}}

\begin{document}

\title{Simulation of quantum physics with Tensor Processing Units:\\
brute-force computation of ground states and time evolution}

\author{Markus Hauru}
\affiliation{Sandbox@Alphabet, Mountain View, CA 94043, USA}

\author{Alan Morningstar}
\affiliation{Department of Physics, Princeton University, Princeton, NJ 08544, USA}
\affiliation{Sandbox@Alphabet, Mountain View, CA 94043, USA}

\author{Jackson Beall}
\affiliation{Sandbox@Alphabet, Mountain View, CA 94043, USA}

\author{Martin Ganahl}
\affiliation{Sandbox@Alphabet, Mountain View, CA 94043, USA}

\author{Adam Lewis}
\affiliation{Sandbox@Alphabet, Mountain View, CA 94043, USA}

\author{Guifre Vidal}
\affiliation{Sandbox@Alphabet, Mountain View, CA 94043, USA}

\date{\today}

\begin{abstract}
Tensor Processing Units (TPUs) were developed by Google exclusively to support large-scale machine learning tasks. TPUs can, however, also be used to accelerate and scale up other computationally demanding tasks. In this paper we repurpose TPUs for the challenging problem of simulating quantum spin systems. Consider a lattice model made of $N$ spin-$\frac{1}{2}$ quantum spins, or qubits, with a Hamiltonian $H = \sum_i h_i$ that is a sum of local terms $h_i$ and a wavefunction $\ket{\Psi}$ consisting of $2^N$ complex amplitudes. We demonstrate the usage of TPUs for both (i) computing the ground state $\ket{\Psi_{\gs}}$ of the Hamiltonian $H$, and (ii) simulating the time evolution $\ket{\Psi(t)}=e^{-itH}\ket{\Psi(0)}$ generated by this Hamiltonian starting from some initial state $\ket{\Psi(0)}$. The bottleneck of the above tasks is computing the product $H \ket{\Psi}$, which can be implemented with remarkable efficiency utilising the native capabilities of TPUs. 
With a TPU v3 pod, with 2048 cores, we simulate wavefunctions $\ket{\Psi}$ of up to $N=38$ qubits. The dedicated matrix multiplication units (MXUs), the high bandwidth memory (HBM) on each core, and the fast inter-core interconnects (ICIs) together provide performance far beyond the capabilities of general purpose processors.
\end{abstract}

\maketitle


The study of quantum many-body phenomena in strongly correlated systems is among the most challenging computational tasks in modern physics.
Describing the quantum mechanical wavefunction of a many-body system, say of $N$ interacting quantum spins, requires computational resources that scale exponentially with the system size $N$.
Sophisticated numerical approaches have been devised over the years to tackle such problems.
For instance, Quantum Monte Carlo (QMC) methods bypass storing the full wavefunction by instead sampling over statistically significant configurations~\cite{qmc_evertz,qmc_prokofev,qmc_syljuasen}, whereas tensor network (TN) algorithms exploit the structure of entanglement in ground states of local Hamiltonians to obtain an efficient quasi-exact description~\cite{white_density_1992,Vidal2004,Verstraete:2004cf,vidal_class_2008,levin_tensor_2007,haegeman_timedependent_2011}.
These highly successful methods are customarily used to study emergent quantum phenomena at scale. They have, however, a restricted range of applicability. For instance, none of them can correctly simulate a long Hamiltonian evolution: QMC due to the sign problem, TNs due to the build-up of entanglement over time. In such situations, a brute-force computation requiring exponentially many resources is still today the only reliable approach. A brute-force computation, free of statistical and/or truncation errors, is also useful even in regimes where QMC and TN methods are expected to work, e.g. to benchmark their performance.

Google's Tensor Processing Units (TPUs) are \textit{application-specific integrated circuits} (ASICs) exclusively designed for accelerating training and inference of machine learning models at scale~\cite{TPUinfo,jouppi2017datacenter}. Here we consider the third generation (v3) of TPUs. We can think of a TPU v3 \textit{pod}, with 2048 cores, 100+ petaFLOPS (in half precision) and 32 TB of high bandwidth memory (HBM), as a special-purpose supercomputer. That is, instead of being designed for general purpose high performance computing (HPC) tasks, a TPU pod is a supercomputer optimized to excel at a class of specialized workloads required for machine learning. Nevertheless, one can still inquire whether TPUs' acceleration and scalability may be repurposed for other computational tasks ~\cite{Belletti-Anderson2020, Wang-Anderson2021, Pan-Mishra2021, Lu-Ma2020, TPUFFT1, TPUFFT2, huot2019highresolution, tpu_algebra, tpu_circuit, 
tpu_floquet, tpu_qhardware, tpu_qchem, tpu_spinLED, tpu_DMRG, tpu_Z2field}, in analogy with how Graphic Processing Units (GPUs), originally conceived to accelerate the rendering of 3D graphics in gaming, have extended to a much broader range of applications, including general purpose HPC and artificial intelligence.

\begin{figure}[tbp]
\centering
\includegraphics[width=1.0\linewidth]{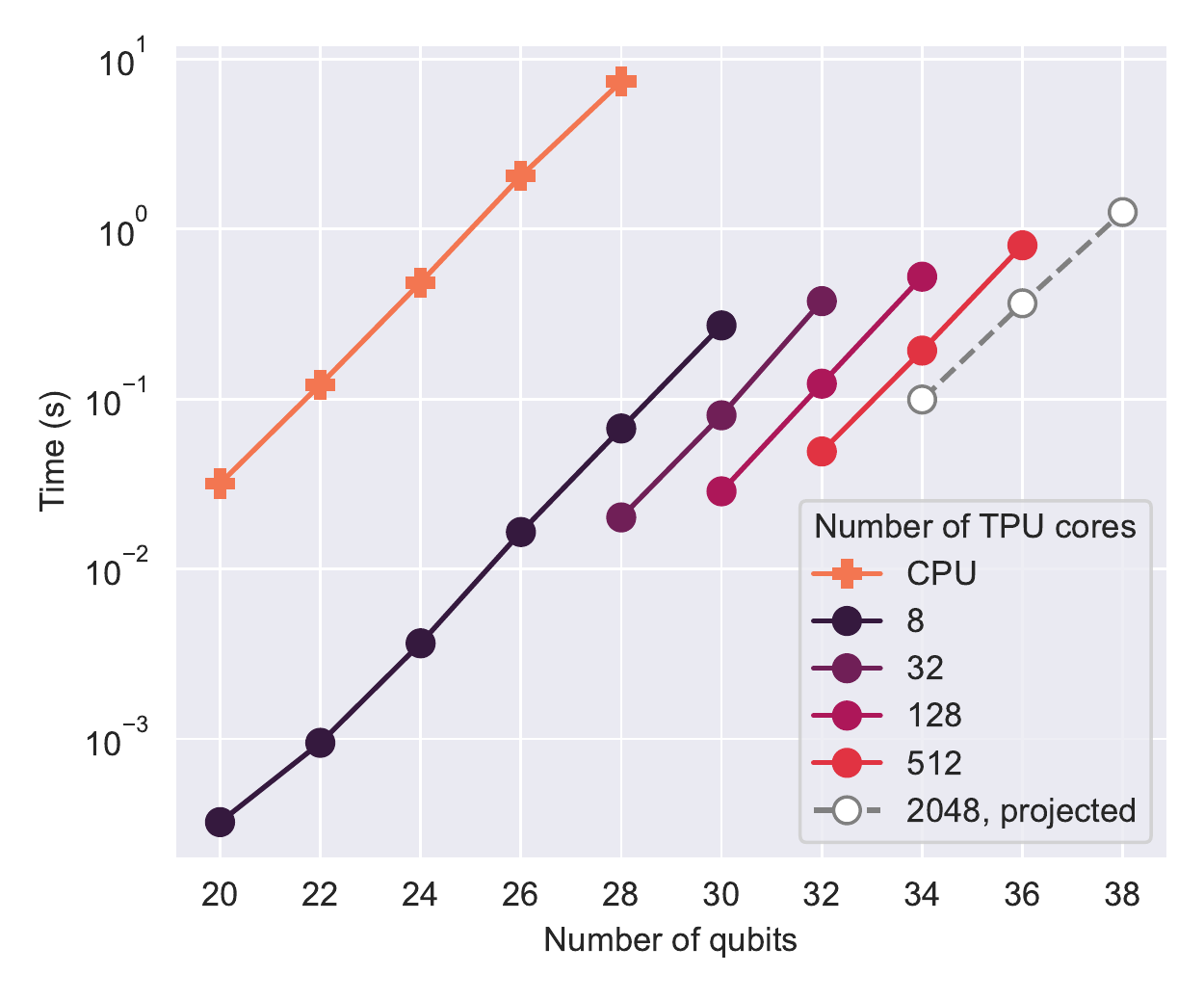}
\caption{%
\label{fig:benchmark}
Compute time for the update $\ket{\Psi}\mapsto H\ket{\Psi}$ for a 1D, nearest neighbor Hamiltonian such as that in Eq. \ref{eq:xxz_hamiltonian}, as a function of the number of qubits. Different colors represent different numbers of TPU cores, ranging from a single board (8 cores) to a full pod (2048 cores). Times are averaged over a large number of repetitions.
Results for 2048 cores are extrapolated estimates, due to temporary resource constraints.
CPU results were run on an 8-core 2.3GHz Intel Xeon workstation, using a Numba-based~\cite{numba} implementation. They represent a typical workstation, rather than cutting-edge high-performance computing hardware.
}
\end{figure}

In this paper we explain how to use TPUs for brute-force simulations of quantum many-body physics. In short, the wavefunction $\ket{\Psi}$ of $N$ spin-$\frac{1}{2}$ quantum spins, or qubits, is distributed over the available HBM and then updated under the action of a local Hamiltonian $H$, 
\begin{equation} \label{eq:HPsi}
    \ket{\Psi} \mapsto \ket{\Psi'} = H \ket{\Psi},
\end{equation}
which requires both large matrix multiplications on each TPU core and repeatedly re-distributing the wavefunction over the cores. Matrix multiplications and wavefunction re-distribution are handled by the TPU's remarkably fast \textit{matrix multiply unit}s (MXUs) and \text{inter-core interconnects} (ICIs), respectively. Fig. \ref{fig:benchmark} shows typical update times
for up to $N=38$ qubits.

The ability to update the wavefunction $\ket{\Psi}$ according to Eq. \eqref{eq:HPsi} can then be used as the basis for a number of computations, which we illustrate for quantum spin chain Hamiltonians. First we obtain the ground state energy $E_{\gs}$ and ground state wavefunction $\ket{\Psi_{\gs}}$ of $H$,
\begin{equation} \label{eq:ground_state}
    H\ket{\Psi_{\gs}} = E_{\gs} \ket{\Psi_{\gs}},~~~~~~~E_{\gs} = \min_{\ket{\Psi}}~\frac{\bra{\Psi}H\ket{\Psi}}{\braket{\Psi|\Psi}}.
\end{equation}
We then use the update \eqref{eq:HPsi} to simulate time evolution according to $H$,
\begin{equation} \label{eq:time_evolution}
    \ket{\Psi(t)} = e^{-itH}\ket{\Psi(0)}, 
\end{equation}
starting from some initial state $\ket{\Psi(0)}$. From the wavefunction $\ket{\Psi}$, one can compute all sorts of derived properties, such as correlation functions or entanglement measures, which we also illustrate. 

\textbf{Tensor Processing Units.---} TPUs come in boards. Each board holds 8 TPU cores controlled by a CPU host. Multiple boards can be joined together to form larger units, up to a so-called a pod, which for third generation TPUs consist of 2048 cores (that is, 256 boards). Each TPU v3 core has 16 GB of high bandwidth memory (HBM) attached to it, which amounts to 128 GB of HBM per board, or 32 TB per pod.
Each TPU v3 core also has two matrix multiplication units (MXUs), which together deliver 52.5 teraFLOPS (floating-point operations per second) of matrix multiplication performance in half precision. MXUs natively perform matrix products by truncating single precision inputs to a half precision format called bfloat16, and accumulating the result again in single precision. Single precision matrix products are then achieved by composing six passes through the MXU (so the above FLOP counts can be divided by six to yield the equivalent single precision FLOPS). Finally, double precision is also available through software emulation with significant additional time and memory overheads \cite{Note_Physics_1}.


A key aspect of TPUs, both in single-board and multi-board configurations, is that all the cores are directly connected to nearest neighbors in a toroidal mesh using fast ICIs and can therefore communicate without going through their CPU host(s). As a result, we observe multi-core performance scaling nearly linearly in the number of cores (e.g. 420 teraFLOPS per board, 100+ petaFLOPS per pod) even in tasks requiring significant communication between cores. 

Code to run on TPUs can be written in several different frameworks, but we use the Python library Jax, which interfaces with the XLA just-in-time compiler~\cite{jax, XLA}. Parallelism between cores follows the single instruction multiple data (SIMD) paradigm. The HBM stores arrays in chunks of $8 \times 128$, and the MXUs natively perform matrix products of $128 \times 128$ matrices. Care must be taken when writing code for TPUs to make sure arrays come in multiples of these sizes, to maintain high performance.

\textbf{Wavefunction distribution.---} Consider the wavefunction $\ket{\Psi}$ of $N$ spin-$\frac{1}{2}$ quantum spins, or $N$ qubits, 
characterized by the $2^{N}$ complex amplitudes $\Psi_{b_1 b_2 \cdots b_N} = \langle b_1 b_2 \cdots b_N|\Psi \rangle$. Each amplitude is indexed by a bit string $(b_1, \dots, b_N)$, where $b_i \in \{0, 1\}$,  and stored here as a pair of single precision floating-point numbers for the real and imaginary parts. Assume that $2^{N_g}$ TPU cores are available. [For instance, a board has $2^3 = 8$ cores and thus $N_g=3$, whereas a full pod has $2^{11}=2048$ cores, or $N_g=11$.] We divide the $N$ qubits into two groups: the first $N_g$ qubits and the remaining $N_l = N - N_g$ qubits, which are called \textit{global} and \textit{local} qubits, respectively. We also decompose each bit-string $(b_1, \cdots, b_N)$ into global and local pieces $(b_1,\cdots, b_{N_g})$ and $(b_{N_g+1}, \cdots, b_{N})$. We then break the $2^N$-element array 
$\Psi_{b_1 \cdots b_{N_g}b_{N_g+1} \cdots b_N}$ into $2^{N_g}$ sub-arrays of $2^{N_l}$ elements each, corresponding to the components with constant global bit-string $(b_1,\cdots, b_{N_g})$, and store each sub-array locally in the HBM of a single TPU core. For instance, for a TPU board ($N_g=3$) we would have 8 sub-arrays distributed over 8 cores, labelled $\#0$ to $\#7$, as follows:

\begin{align*}
\includegraphics[width=0.9\linewidth]{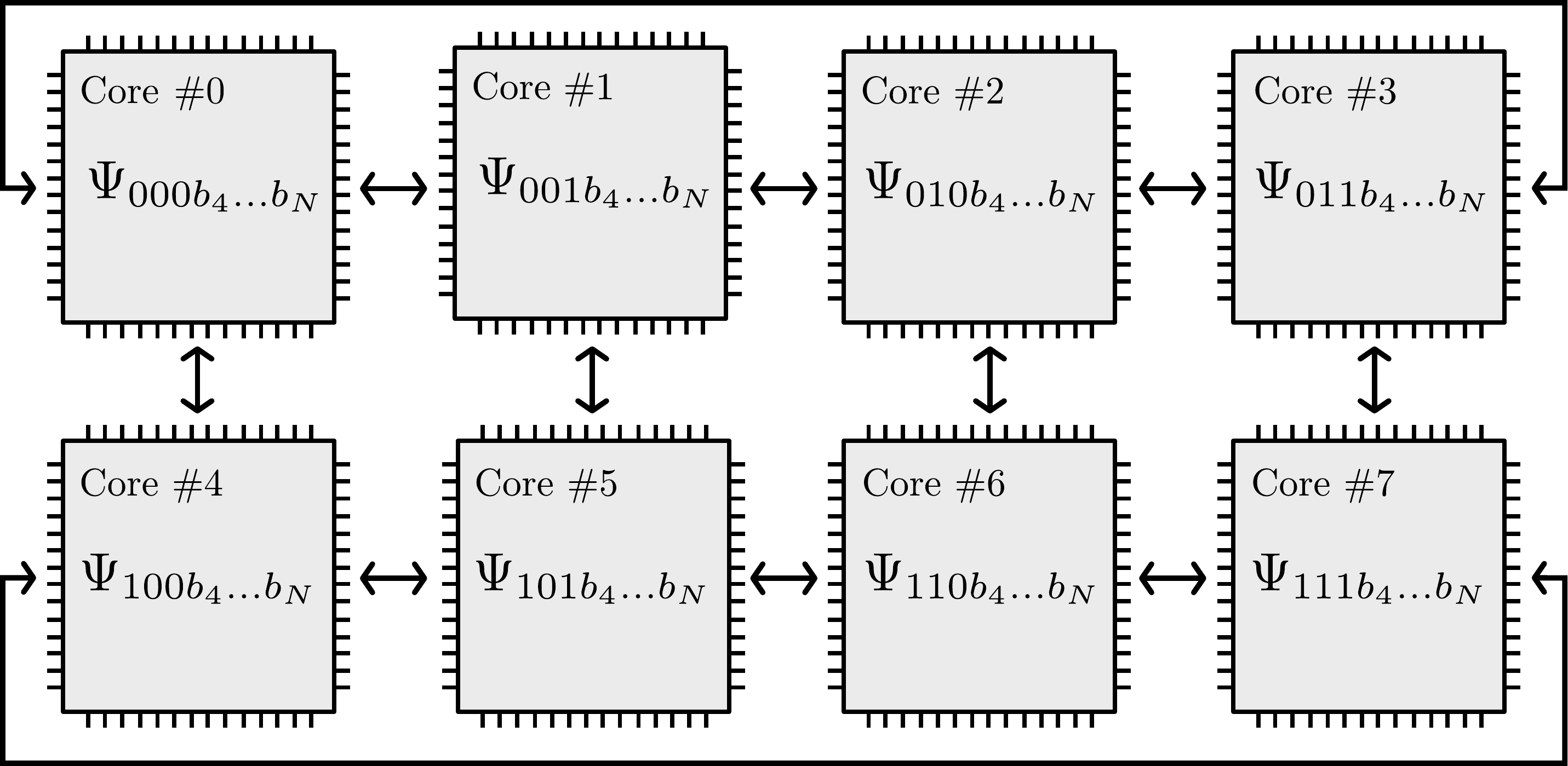}
\end{align*}
We denote the distributed wavefunction with the following diagram, where each line corresponds to a qubit (red lines on the left correspond to global qubits):
\begin{align*}
\includegraphics[width=0.9\linewidth]{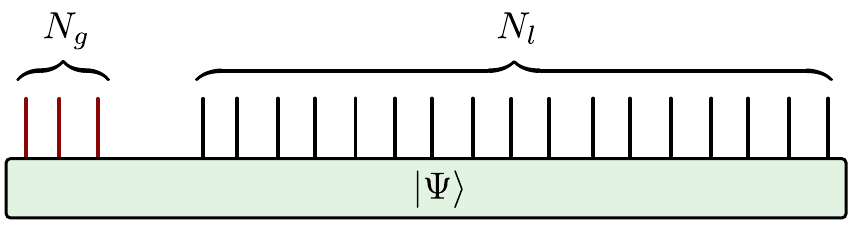}
\end{align*}
TPU memory considerations require that we store each local sub-array as an array with two or more dimensions, where the last two dimensions are multiples of 8 and 128 respectively, see App.~\ref{app:distribution} for more details.

\textbf{Wavefunction update.---} Consider now a local Hamiltonian $H = \sum_{i=1}^{P} h_i$ that decomposes as a sum of $P$ local terms $h_i$, each acting on $7$ qubits. We focus on 7-qubit terms because they fit the memory constraints of TPUs, see App.~\ref{app:HPsi} for how to turn a more generic local Hamiltonian into this form. In order to apply $H$ on $\ket{\Psi}$ as in Eq. \eqref{eq:HPsi}, we compute a sequence of products $h_i \ket{\Psi}$ for $i=1,\cdots,P$, which we accumulate in wavefunctions $\ket{\Psi^{(i)}}$,
\begin{equation}
    \ket{\Psi^{(i)}} = \ket{\Psi^{(i-1)}} + h_{i} \ket{\Psi},~~~~i=1,\cdots, P, 
\end{equation}
where $\ket{\Psi^{(0)}} =0 $ is the null vector and $\ket{\Psi^{(P)}} = H \ket{\Psi}$ contains the final result.
Each Hamiltonian term $h_i$ is represented by a $2^7\times 2^7=128\times 128$ matrix and broadcast to each TPU core. Assume first that the term $h_i$ acts on the last 7 local qubits. We then simply multiply the distributed array for $\Psi$, regarded as a $2^{N-7}\times 2^{7}$ matrix, with the $2^7 \times 2^7$ matrix $h_i$, represented as:
\begin{align*}
\includegraphics[width=0.9\linewidth]{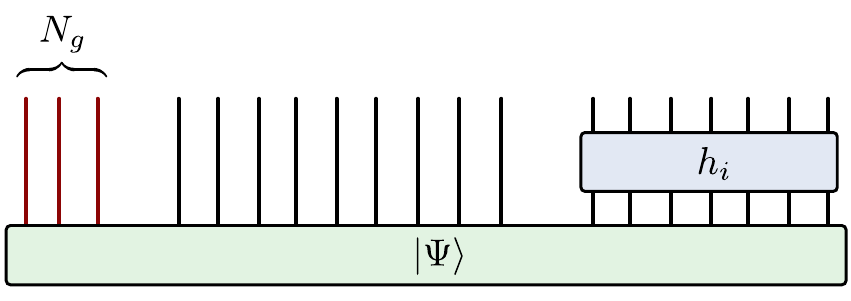}
\end{align*}
This product is performed for each sub-array locally on each TPU core, utilizing the MXUs. When $h_i$ acts on other local qubits, we can choose to permute their order so that $h_i$ effectively acts on the last 7 local qubits. This requires reshaping and/or transposing the local sub-array of $\Psi$ on each core, while paying due attention to the restrictions on array shapes mentioned above (see also App.~\ref{app:distribution}). All the above operations are executed in parallel and without inter-core communication, with each core manipulating their local sub-array according to an identical set of instructions. 

Finally, when the term $h_i$ acts on at least one global qubit, we first re-distribute the wavefunction $\ket{\Psi}$, by swapping (possibly a subset of) the global qubits with an equivalent number of local qubits, in such a way that the Hamiltonian term $h_i$ ends up acting on local qubits, at which point we can compute $h_i \ket{\Psi}$ as described above. For instance, in a single-board (i.e. 8-core) set-up, with $N_g=3$, we may want to swap the three global qubits with the first three local qubits, represented by
\begin{align*}
\includegraphics[width=0.9\linewidth]{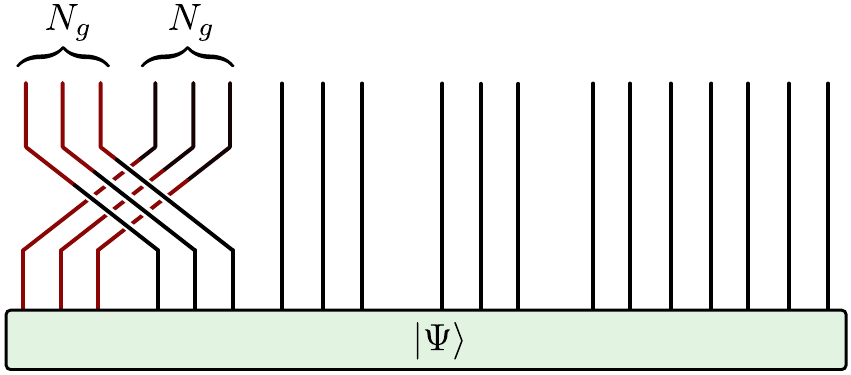}
\end{align*}
The re-distribution of $\ket{\Psi}$, executed remarkably fast thanks to the ICIs, requires substantial communication between the cores. Fig. \ref{fig:benchmark} shows the total time required for the update $\ket{\Psi} \mapsto H\ket{\Psi}$ for an example Hamiltonian. Using a full TPU v3 pod, the wavefunction for $N=38$ qubits is updated in about 1 second.

\begin{figure}[tbp]
\centering
\includegraphics[width=1.0\linewidth]{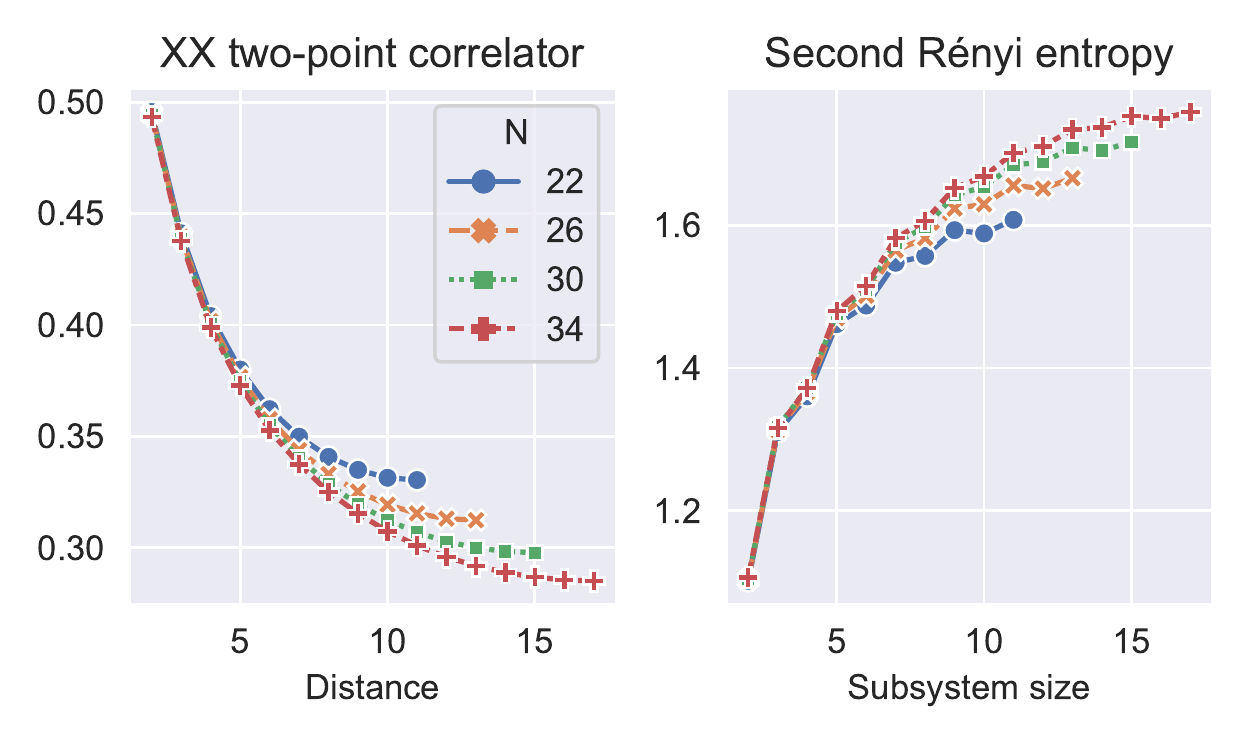}
\caption{%
\label{fig:groundstate_both}
Properties of the ground state of the 1D XXZ model in Eq. \eqref{eq:xxz_hamiltonian} with $J = -1$, $\Delta = \frac{1}{2}$.
Left:
The $\langle X_1 X_n \rangle - \langle X_1 \rangle \langle X_n \rangle$ connected two-point correlator as a function of the distance $n$.
Right:
Second Rényi entropy $S_2$ of a subsystem as a function of subsystem size.
The different lines are for different system sizes.
}
\end{figure}

\textbf{Computation of ground states.---} To illustrate the computation of a ground state $\ket{\Psi_{\gs}}$, Eq. \eqref{eq:ground_state}, we apply the Lanczos algorithm, based on the update $\ket{\Psi}\mapsto H\ket{\Psi}$ as described in App.~\ref{app:lanczos}, to the 1D XXZ Hamiltonian,
\begin{equation}
    \label{eq:xxz_hamiltonian}
H = J \sum_{i=1}^{N} (X_i X_{i+1} + Y_i Y_{i+1} + \Delta Z_i Z_{i+1}),
\end{equation}
where $X$, $Y$, and $Z$ are the Pauli matrices and site $N$ is identified with site 0 (periodic boundary conditions).  For the couplings $J = -1$ and $\Delta = \frac{1}{2}$, this model is known to be at a quantum critical point, in the universality class of a bosonic conformal field theory with unit central charge ~\cite{henkel_conformal_1999}. Our current implementation, which handles generic 1D local Hamiltonians, does not exploit any of the several symmetries of the XXZ model, e.g. translation invariance or internal U(1) symmetry. The 2-qubit terms in the above Hamiltonian are merged together into 7-qubit terms (see App. \ref{app:HPsi}) to optimize the use of the MXUs. For this computation we use at most 128 TPU v3 cores ($N_g=7$) to reach up to $N=34$ qubits, with the total time for the computation being around 12 minutes.

Importantly, from the distributed array for $\ket{\Psi}$ we can extract a large variety of quantities, including e.g. wavefunction overlaps $\braket{\Psi_1|\Psi_2}$, expectation value of observables such as the energy $\braket{\Psi|H|\Psi}$, correlation functions and entanglement measures. This requires additional operations that can also be easily implemented on TPUs. The reduced density matrix $\rho_A = \Tr_{\bar{A}} |\Psi \rangle \langle \Psi |$ over a subset $A$ of qubits (where the trace is over the rest of the qubits, denoted $\bar{A}$), and derived quantities such as von Neumann and Rényi entropies can also be computed for subsystems of up to $N/2$ qubits. Handling them requires distributed linear algebra methods on TPUs, discussed in Ref.~\onlinecite{tpu_algebra}. As an example, Fig.~\ref{fig:groundstate_both} shows the connected two-point correlator $\bra{\Psi} X_1 X_n \ket{\Psi} - \bra{\Psi} X_1 \ket{\Psi} \bra{\Psi} X_n \ket{\Psi}$ as a function of the distance $n$ between spins, as well as the second Rényi entropy $S_2(\rho_A) = -\log_2 \Tr[(\rho_A)^2]$, where $\rho_A$ corresponds to a subsystem $A$ of contiguous spins.

\textbf{Simulation of time evolution.---} Our second application is the simulation of a time evolution $\ket{\Psi(t)} = e^{-itH}\ket{\Psi(0)}$ that generates large amounts of entanglement.
We define a suitably small time interval $\delta t$ such that $q \equiv t/\delta t$ is an integer, write the time evolution operator for time $t$, $\exp(-i t H)$, as the $q$-fold product of the time operator for $\delta t$, $\exp(-i t H) = \prod_{j=1}^q \exp(-i \, \delta t \, H)$,
and then approximate each operator $\exp(-i \, \delta t \, H)$ as a sequence of six terms of the form $(1-i a_n \delta t H)$,
\begin{eqnarray}
    &&\exp(-i \, \delta t \, H) = (1-i a_6 \delta t H) \times \cdots~~~~~~~~\nonumber\\
    &&~~~~~\cdots \times (1-i a_2 \delta t H)\times (1-i a_1 \delta t H) + O(\delta t^7),~~
\end{eqnarray}
where the coefficients $a_n$ are solved for numerically, see App.~\ref{app:time-evolution-expansion}. The dominant term neglected above is  $\frac{H^7 \delta t^7}{7!}$. Each update $\ket{\Psi} \mapsto (1 -i \, a_n \, \delta t \, H) \ket{\Psi}$ requires computing $\ket{\Psi} \mapsto H\ket{\Psi}$ and is implemented while keeping only a few copies of $\ket{\Psi}$ in memory simultaneously.

For this demonstration we choose a 1D local Hamiltonian of the form $H = \sum_{i=1}^{N} h_{[i, i+5]}$, where each $h_{[i, i+5]}$ is a Hermitian matrix of size $2^{6} \times 2^{6}$ that operates on qubits $i$ through $i+5$, thus exemplifying the use of $k$-qubit Hamiltonian terms for $k>2$. $H$ is set to have periodic boundaries. The matrix elements of $h_{[i, i+5]}$, which represents a generic non-integrable many-body interaction, are Gaussian distributed with mean zero and standard deviation $\sqrt{6}\; 2^{-5}$ (such that the average Frobenius norm of each term is $\langle \|h_{[i, i+5]}\| \rangle = \sqrt{6}$). The 6-qubit terms $h_{[i, i+5]}$ are still pairwise blocked into 7-qubit terms (see App. \ref{app:HPsi}) in order to optimally utilize the MXUs.

\begin{figure}[tbp]
\centering
\includegraphics[width=1.0\linewidth]{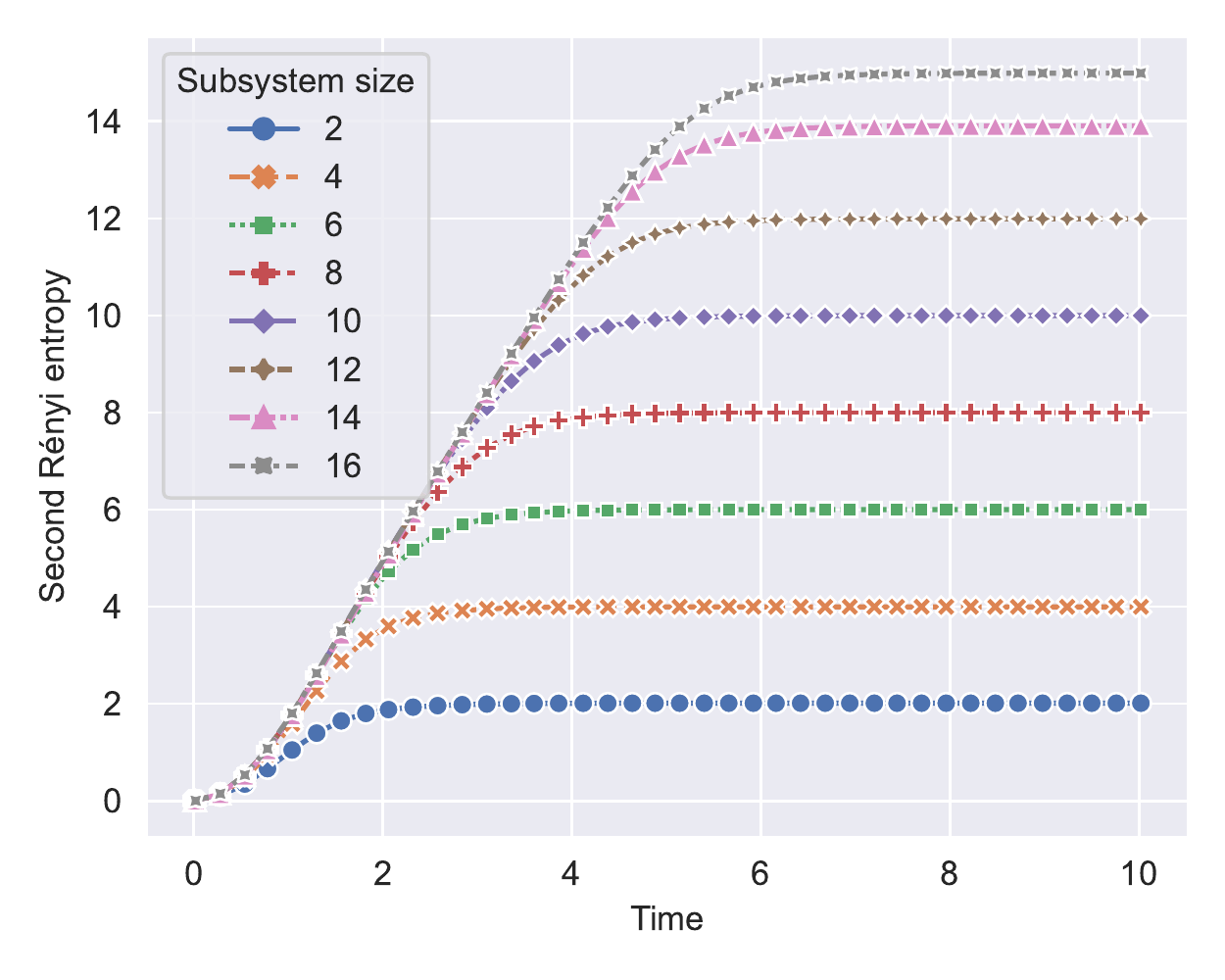}
\caption{%
\label{fig:evolve_entropy}
Second Rényi entropy $S_2$ of various subsystems, as a function of time $t$, during an entangling time evolution $\ket{\Psi(t)}= e^{-itH}\ket{\Psi(0)}$ on $N=32$ qubits, starting from an unentangled product state $\ket{\Psi(0)}$ and according to a 1D random Hamiltonian with 6-qubit terms. For each subsystem size, the entropy initially grows until saturating to a maximum value corresponding to a random wavefunction.
}
\end{figure}

Fig.~\ref{fig:evolve_entropy} shows the second Rényi entropy, as a function of time $t$, of various subsystems for the state $\ket{\Psi(t)} = \exp(-i t H)\ket{\Psi(0)}$, where the initial state $\ket{\Psi(0)}$ has no entanglement, and the system has $N=32$ qubits. On 32 TPU v3 cores ($N_g=5$), simulating this evolution, which includes $q=500$ time steps with $\delta t=0.02$, and evaluation of the entropies, took 44 minutes.
At early times the entropy is seen to grow similarly for all subsystems, regardless of their size. However, in time each subsystem saturates to a constant amount of entropy, dependent only on the size of the subsystem. Specifically, for a subsystem of $M$ qubits this saturation value is roughly $M$ bits of entropy, except for subsystems that cover a large proportion of the total system, such as $M = \frac{N}{2}$, for which it is somewhat less, as expected \cite{Hayden2006}.
For the larger subsystems we see a sustained period of linear growth, which is evidence for ballistic growth of entanglement, as expected for a random Hamiltonian~\cite{KimHuse2013}. The rapid growth of entanglement all the way to saturation makes this time evolution very hard to simulate using tensor network methods such as Matrix Product States (MPS), which rely on the wavefunction being only moderately entangled and thus compressible. Indeed, in this example an MPS description would require a central bond dimension close to $2^{16}=65,536$. In other words, the largest MPS tensors have a similar size as the whole wavefunction ($2^{32}$ complex coefficients), making the MPS approach even more costly than the present brute-force approach.

\textbf{Discussion.---} TPUs, originally designed for machine learning workloads, can be effectively repurposed to accelerate and scale up a number of other computationally demanding tasks 
~\cite{Belletti-Anderson2020, Wang-Anderson2021, Pan-Mishra2021, Lu-Ma2020, TPUFFT1, TPUFFT2, huot2019highresolution, tpu_algebra, tpu_circuit, 
tpu_floquet, tpu_qhardware, tpu_qchem, tpu_spinLED, tpu_DMRG, tpu_Z2field}. In this paper we have demonstrated that TPUs can be used to compute ground states and simulate real time evolution in many-body systems. A very high matrix multiplication throughput on each core (enabled by the MXUs), combined with the ability to directly connect up to thousands of TPU cores together (using the fast ICIs) for a single parallel simulation with dozens of terabytes of HBM allowed us to consider an unusually large number $N$ of qubits at remarkable speed, see e.g. Fig. \ref{fig:benchmark}. This required adjusting to a number of characteristic features of TPUs and their XLA compiler, such as operating under the SIMD paradigm and using specific shapes for the matrices and arrays. Importantly, operating with $k$-qubit Hamiltonian terms for $k<7$ has a computational cost similar to the $k=7$ case, for which we observe peak performance.

Here we have considered local Hamiltonians $H$ in one dimension for concreteness. It is of course also possible to address 2D/3D local Hamiltonians, e.g. for square/cubic lattices, where sizes $6\times 6$,  $4 \times 9$ or $4 \times 3 \times 3$ are easily accessible. To go well beyond these system sizes, one has to give up a brute-force, exact simulation based on storing the full wavefunction, and use instead approximate methods such as tensor network approaches. Tensor network methods can also be significantly accelerated and scaled up with TPUs, allowing e.g. for MPS representations of 100-qubit wavefunctions with unusually large bond dimensions~\cite{tpu_DMRG}, capable of describing larger amounts of entanglement than it is possible on regular hardware.


\begin{acknowledgments}
This research was supported with Cloud TPUs from Google's TPU Research Cloud (TRC).
Sandbox is a team within the Alphabet family of companies, which includes Google, Verily, Waymo, X, and others.
GV is a CIFAR fellow in the Quantum Information Science Program and a Distinguished Visiting Research Chair at Perimeter Institute.
Research at Perimeter Institute is supported by the Government of Canada through the Department of Innovation, Science and Economic Development and by the Province of Ontario through the Ministry of Research, Innovation and Science.
\end{acknowledgments}

\bibliography{main}

\begin{thebibliography}{33}%
\makeatletter
\providecommand \@ifxundefined [1]{%
 \@ifx{#1\undefined}
}%
\providecommand \@ifnum [1]{%
 \ifnum #1\expandafter \@firstoftwo
 \else \expandafter \@secondoftwo
 \fi
}%
\providecommand \@ifx [1]{%
 \ifx #1\expandafter \@firstoftwo
 \else \expandafter \@secondoftwo
 \fi
}%
\providecommand \natexlab [1]{#1}%
\providecommand \enquote  [1]{``#1''}%
\providecommand \bibnamefont  [1]{#1}%
\providecommand \bibfnamefont [1]{#1}%
\providecommand \citenamefont [1]{#1}%
\providecommand \href@noop [0]{\@secondoftwo}%
\providecommand \href [0]{\begingroup \@sanitize@url \@href}%
\providecommand \@href[1]{\@@startlink{#1}\@@href}%
\providecommand \@@href[1]{\endgroup#1\@@endlink}%
\providecommand \@sanitize@url [0]{\catcode `\\12\catcode `\$12\catcode
  `\&12\catcode `\#12\catcode `\^12\catcode `\_12\catcode `\%12\relax}%
\providecommand \@@startlink[1]{}%
\providecommand \@@endlink[0]{}%
\providecommand \url  [0]{\begingroup\@sanitize@url \@url }%
\providecommand \@url [1]{\endgroup\@href {#1}{\urlprefix }}%
\providecommand \urlprefix  [0]{URL }%
\providecommand \Eprint [0]{\href }%
\providecommand \doibase [0]{http://dx.doi.org/}%
\providecommand \selectlanguage [0]{\@gobble}%
\providecommand \bibinfo  [0]{\@secondoftwo}%
\providecommand \bibfield  [0]{\@secondoftwo}%
\providecommand \translation [1]{[#1]}%
\providecommand \BibitemOpen [0]{}%
\providecommand \bibitemStop [0]{}%
\providecommand \bibitemNoStop [0]{.\EOS\space}%
\providecommand \EOS [0]{\spacefactor3000\relax}%
\providecommand \BibitemShut  [1]{\csname bibitem#1\endcsname}%
\let\auto@bib@innerbib\@empty
\bibitem [{\citenamefont {Evertz}(2003)}]{qmc_evertz}%
  \BibitemOpen
  \bibfield  {author} {\bibinfo {author} {\bibfnamefont {H.~G.}\ \bibnamefont
  {Evertz}},\ }\bibfield  {title} {\enquote {\bibinfo {title} {The loop
  algorithm},}\ }\href {\doibase 10.1080/0001873021000049195} {\bibfield
  {journal} {\bibinfo  {journal} {Advances in Physics}\ }\textbf {\bibinfo
  {volume} {52}},\ \bibinfo {pages} {1--66} (\bibinfo {year} {2003})},\ \Eprint
  {http://arxiv.org/abs/cond-mat/9707221} {arXiv:cond-mat/9707221} \BibitemShut
  {NoStop}%
\bibitem [{\citenamefont {{Prokof'ev}}\ \emph {et~al.}(1998)\citenamefont
  {{Prokof'ev}}, \citenamefont {{Svistunov}},\ and\ \citenamefont
  {{Tupitsyn}}}]{qmc_prokofev}%
  \BibitemOpen
  \bibfield  {author} {\bibinfo {author} {\bibfnamefont {N.~V.}\ \bibnamefont
  {{Prokof'ev}}}, \bibinfo {author} {\bibfnamefont {B.~V.}\ \bibnamefont
  {{Svistunov}}}, \ and\ \bibinfo {author} {\bibfnamefont {I.~S.}\ \bibnamefont
  {{Tupitsyn}}},\ }\bibfield  {title} {\enquote {\bibinfo {title} {{Exact,
  complete, and universal continuous-time worldline {Monte Carlo} approach to
  the statistics of discrete quantum systems}},}\ }\href {\doibase
  10.1134/1.558661} {\bibfield  {journal} {\bibinfo  {journal} {Journal of
  Experimental and Theoretical Physics}\ }\textbf {\bibinfo {volume} {87}},\
  \bibinfo {pages} {310--321} (\bibinfo {year} {1998})},\ \Eprint
  {http://arxiv.org/abs/cond-mat/9703200} {arXiv:cond-mat/9703200} \BibitemShut
  {NoStop}%
\bibitem [{\citenamefont {Sylju\aa{}sen}\ and\ \citenamefont
  {Sandvik}(2002)}]{qmc_syljuasen}%
  \BibitemOpen
  \bibfield  {author} {\bibinfo {author} {\bibfnamefont {Olav~F.}\ \bibnamefont
  {Sylju\aa{}sen}}\ and\ \bibinfo {author} {\bibfnamefont {Anders~W.}\
  \bibnamefont {Sandvik}},\ }\bibfield  {title} {\enquote {\bibinfo {title}
  {Quantum {Monte Carlo} with directed loops},}\ }\href {\doibase
  10.1103/PhysRevE.66.046701} {\bibfield  {journal} {\bibinfo  {journal} {Phys.
  Rev. E}\ }\textbf {\bibinfo {volume} {66}},\ \bibinfo {pages} {046701}
  (\bibinfo {year} {2002})},\ \Eprint {http://arxiv.org/abs/cond-mat/0202316}
  {arXiv:cond-mat/0202316} \BibitemShut {NoStop}%
\bibitem [{\citenamefont {{White}}(1992)}]{white_density_1992}%
  \BibitemOpen
  \bibfield  {author} {\bibinfo {author} {\bibfnamefont {Steven~R.}\
  \bibnamefont {{White}}},\ }\bibfield  {title} {\enquote {\bibinfo {title}
  {Density matrix formulation for quantum renormalization groups},}\ }\href
  {http://link.aps.org/doi/10.1103/PhysRevLett.69.2863} {\bibfield  {journal}
  {\bibinfo  {journal} {Phys. Rev. Lett.}\ }\textbf {\bibinfo {volume} {69}},\
  \bibinfo {pages} {2863--2866} (\bibinfo {year} {1992})}\BibitemShut {NoStop}%
\bibitem [{\citenamefont {Vidal}(2004)}]{Vidal2004}%
  \BibitemOpen
  \bibfield  {author} {\bibinfo {author} {\bibfnamefont {Guifr\'e}\
  \bibnamefont {Vidal}},\ }\bibfield  {title} {\enquote {\bibinfo {title}
  {Efficient simulation of one-dimensional quantum many-body systems},}\ }\href
  {\doibase 10.1103/PhysRevLett.93.040502} {\bibfield  {journal} {\bibinfo
  {journal} {Phys. Rev. Lett.}\ }\textbf {\bibinfo {volume} {93}},\ \bibinfo
  {pages} {040502} (\bibinfo {year} {2004})}\BibitemShut {NoStop}%
\bibitem [{\citenamefont {Verstraete}\ and\ \citenamefont
  {Cirac}(2004)}]{Verstraete:2004cf}%
  \BibitemOpen
  \bibfield  {author} {\bibinfo {author} {\bibfnamefont {F.}~\bibnamefont
  {Verstraete}}\ and\ \bibinfo {author} {\bibfnamefont {J.~I.}\ \bibnamefont
  {Cirac}},\ }\bibfield  {title} {\enquote {\bibinfo {title} {{Renormalization
  algorithms for quantum-many body systems in two and higher dimensions}},}\
  }\href@noop {} {\  (\bibinfo {year} {2004})},\ \Eprint
  {http://arxiv.org/abs/cond-mat/0407066} {arXiv:cond-mat/0407066} \BibitemShut
  {NoStop}%
\bibitem [{\citenamefont {{Vidal}}(2008)}]{vidal_class_2008}%
  \BibitemOpen
  \bibfield  {author} {\bibinfo {author} {\bibfnamefont {G.}~\bibnamefont
  {{Vidal}}},\ }\bibfield  {title} {\enquote {\bibinfo {title} {{A} class of
  quantum many-body states that can be efficiently simulated},}\ }\href
  {\doibase 10.1103/PhysRevLett.101.110501} {\bibfield  {journal} {\bibinfo
  {journal} {Phys. Rev. Lett.}\ }\textbf {\bibinfo {volume} {101}},\ \bibinfo
  {pages} {110501} (\bibinfo {year} {2008})},\ \Eprint
  {http://arxiv.org/abs/quant-ph/0610099} {arXiv:quant-ph/0610099} \BibitemShut
  {NoStop}%
\bibitem [{\citenamefont {{Levin}}\ and\ \citenamefont
  {{Nave}}(2007)}]{levin_tensor_2007}%
  \BibitemOpen
  \bibfield  {author} {\bibinfo {author} {\bibfnamefont {Michael}\ \bibnamefont
  {{Levin}}}\ and\ \bibinfo {author} {\bibfnamefont {Cody~P.}\ \bibnamefont
  {{Nave}}},\ }\bibfield  {title} {\enquote {\bibinfo {title} {Tensor
  renormalization group approach to {2D} classical lattice models},}\ }\href
  {http://arxiv.org/abs/cond-mat/0611687} {\bibfield  {journal} {\bibinfo
  {journal} {Phys. Rev. Lett.}\ }\textbf {\bibinfo {volume} {99}},\ \bibinfo
  {pages} {120601} (\bibinfo {year} {2007})},\ \Eprint
  {http://arxiv.org/abs/cond-mat/0611687} {arXiv:cond-mat/0611687} \BibitemShut
  {NoStop}%
\bibitem [{\citenamefont {{Haegeman}}\ \emph {et~al.}(2011)\citenamefont
  {{Haegeman}}, \citenamefont {{Cirac}}, \citenamefont {{Osborne}},
  \citenamefont {{Pizorn}}, \citenamefont {{Verschelde}},\ and\ \citenamefont
  {{Verstraete}}}]{haegeman_timedependent_2011}%
  \BibitemOpen
  \bibfield  {author} {\bibinfo {author} {\bibfnamefont {Jutho}\ \bibnamefont
  {{Haegeman}}}, \bibinfo {author} {\bibfnamefont {J.~Ignacio}\ \bibnamefont
  {{Cirac}}}, \bibinfo {author} {\bibfnamefont {Tobias~J.}\ \bibnamefont
  {{Osborne}}}, \bibinfo {author} {\bibfnamefont {Iztok}\ \bibnamefont
  {{Pizorn}}}, \bibinfo {author} {\bibfnamefont {Henri}\ \bibnamefont
  {{Verschelde}}}, \ and\ \bibinfo {author} {\bibfnamefont {Frank}\
  \bibnamefont {{Verstraete}}},\ }\bibfield  {title} {\enquote {\bibinfo
  {title} {Time-dependent variational principle for quantum lattices},}\ }\href
  {http://arxiv.org/abs/1103.0936} {\bibfield  {journal} {\bibinfo  {journal}
  {Phys. Rev. Lett.}\ }\textbf {\bibinfo {volume} {107}} (\bibinfo {year}
  {2011})},\ \Eprint {http://arxiv.org/abs/1103.0936} {arXiv:1103.0936}
  \BibitemShut {NoStop}%
\bibitem [{\citenamefont {Jouppi}\ \emph {et~al.}(2020)\citenamefont {Jouppi},
  \citenamefont {Yoon}, \citenamefont {Kurian}, \citenamefont {Li},
  \citenamefont {Patil}, \citenamefont {Laudon}, \citenamefont {Young},\ and\
  \citenamefont {Patterson}}]{TPUinfo}%
  \BibitemOpen
  \bibfield  {author} {\bibinfo {author} {\bibfnamefont {Norman}\ \bibnamefont
  {Jouppi}}, \bibinfo {author} {\bibfnamefont {Doe}\ \bibnamefont {Yoon}},
  \bibinfo {author} {\bibfnamefont {George}\ \bibnamefont {Kurian}}, \bibinfo
  {author} {\bibfnamefont {Sheng}\ \bibnamefont {Li}}, \bibinfo {author}
  {\bibfnamefont {Nishant}\ \bibnamefont {Patil}}, \bibinfo {author}
  {\bibfnamefont {James}\ \bibnamefont {Laudon}}, \bibinfo {author}
  {\bibfnamefont {Cliff}\ \bibnamefont {Young}}, \ and\ \bibinfo {author}
  {\bibfnamefont {David}\ \bibnamefont {Patterson}},\ }\bibfield  {title}
  {\enquote {\bibinfo {title} {A domain-specific supercomputer for training
  deep neural networks},}\ }\href {\doibase 10.1145/3360307} {\bibfield
  {journal} {\bibinfo  {journal} {Communications of the ACM}\ }\textbf
  {\bibinfo {volume} {63}},\ \bibinfo {pages} {67--78} (\bibinfo {year}
  {2020})}\BibitemShut {NoStop}%
\bibitem [{\citenamefont {Jouppi}\ \emph {et~al.}(2017)\citenamefont {Jouppi},
  \citenamefont {Young}, \citenamefont {Patil}, \citenamefont {Patterson},
  \citenamefont {Agrawal}, \citenamefont {Bajwa}, \citenamefont {Bates},
  \citenamefont {Bhatia}, \citenamefont {Boden}, \citenamefont {Borchers} \emph
  {et~al.}}]{jouppi2017datacenter}%
  \BibitemOpen
  \bibfield  {author} {\bibinfo {author} {\bibfnamefont {Norman~P.}\
  \bibnamefont {Jouppi}}, \bibinfo {author} {\bibfnamefont {Cliff}\
  \bibnamefont {Young}}, \bibinfo {author} {\bibfnamefont {Nishant}\
  \bibnamefont {Patil}}, \bibinfo {author} {\bibfnamefont {David}\ \bibnamefont
  {Patterson}}, \bibinfo {author} {\bibfnamefont {Gaurav}\ \bibnamefont
  {Agrawal}}, \bibinfo {author} {\bibfnamefont {Raminder}\ \bibnamefont
  {Bajwa}}, \bibinfo {author} {\bibfnamefont {Sarah}\ \bibnamefont {Bates}},
  \bibinfo {author} {\bibfnamefont {Suresh}\ \bibnamefont {Bhatia}}, \bibinfo
  {author} {\bibfnamefont {Nan}\ \bibnamefont {Boden}}, \bibinfo {author}
  {\bibfnamefont {Al}~\bibnamefont {Borchers}},  \emph {et~al.},\ }\bibfield
  {title} {\enquote {\bibinfo {title} {In-datacenter performance analysis of a
  tensor processing unit},}\ }in\ \href {\doibase 10.1145/3079856.3080246}
  {\emph {\bibinfo {booktitle} {{Proceedings of the 44th Annual International
  Symposium on Computer Architecture}}}},\ \bibinfo {series and number} {ISCA
  '17}\ (\bibinfo  {publisher} {Association for Computing Machinery},\ \bibinfo
  {address} {New York, NY, USA},\ \bibinfo {year} {2017})\ p.\ \bibinfo {pages}
  {1–12}\BibitemShut {NoStop}%
\bibitem [{\citenamefont {Belletti}\ \emph {et~al.}(2020)\citenamefont
  {Belletti}, \citenamefont {King}, \citenamefont {Yang}, \citenamefont
  {Nelet}, \citenamefont {Shafi}, \citenamefont {Chen},\ and\ \citenamefont
  {Anderson}}]{Belletti-Anderson2020}%
  \BibitemOpen
  \bibfield  {author} {\bibinfo {author} {\bibfnamefont {Francois}\
  \bibnamefont {Belletti}}, \bibinfo {author} {\bibfnamefont {Davis}\
  \bibnamefont {King}}, \bibinfo {author} {\bibfnamefont {Kun}\ \bibnamefont
  {Yang}}, \bibinfo {author} {\bibfnamefont {Roland}\ \bibnamefont {Nelet}},
  \bibinfo {author} {\bibfnamefont {Yusef}\ \bibnamefont {Shafi}}, \bibinfo
  {author} {\bibfnamefont {Yi-Fan}\ \bibnamefont {Chen}}, \ and\ \bibinfo
  {author} {\bibfnamefont {John}\ \bibnamefont {Anderson}},\ }\href@noop {}
  {\enquote {\bibinfo {title} {Tensor processing units for financial {Monte
  Carlo}},}\ } (\bibinfo {year} {2020}),\ \Eprint
  {http://arxiv.org/abs/1906.02818} {arXiv:1906.02818 [cs.DC]} \BibitemShut
  {NoStop}%
\bibitem [{\citenamefont {Wang}\ \emph {et~al.}(2021)\citenamefont {Wang},
  \citenamefont {Ihme}, \citenamefont {Chen},\ and\ \citenamefont
  {Anderson}}]{Wang-Anderson2021}%
  \BibitemOpen
  \bibfield  {author} {\bibinfo {author} {\bibfnamefont {Qing}\ \bibnamefont
  {Wang}}, \bibinfo {author} {\bibfnamefont {Matthias}\ \bibnamefont {Ihme}},
  \bibinfo {author} {\bibfnamefont {Yi-Fan}\ \bibnamefont {Chen}}, \ and\
  \bibinfo {author} {\bibfnamefont {John}\ \bibnamefont {Anderson}},\
  }\href@noop {} {\enquote {\bibinfo {title} {A tensorflow simulation framework
  for scientific computing of fluid flows on tensor processing units},}\ }
  (\bibinfo {year} {2021}),\ \Eprint {http://arxiv.org/abs/2108.11076}
  {arXiv:2108.11076 [physics.comp-ph]} \BibitemShut {NoStop}%
\bibitem [{\citenamefont {Pan}\ and\ \citenamefont
  {Mishra}(2021)}]{Pan-Mishra2021}%
  \BibitemOpen
  \bibfield  {author} {\bibinfo {author} {\bibfnamefont {Zhixin}\ \bibnamefont
  {Pan}}\ and\ \bibinfo {author} {\bibfnamefont {Prabhat}\ \bibnamefont
  {Mishra}},\ }\href@noop {} {\enquote {\bibinfo {title} {Hardware acceleration
  of explainable machine learning using tensor processing units},}\ } (\bibinfo
  {year} {2021}),\ \Eprint {http://arxiv.org/abs/2103.11927} {arXiv:2103.11927
  [cs.LG]} \BibitemShut {NoStop}%
\bibitem [{\citenamefont {Lu}\ \emph {et~al.}(2020{\natexlab{a}})\citenamefont
  {Lu}, \citenamefont {Marin}, \citenamefont {Zhuo}, \citenamefont {Chen},\
  and\ \citenamefont {Ma}}]{Lu-Ma2020}%
  \BibitemOpen
  \bibfield  {author} {\bibinfo {author} {\bibfnamefont {Tianjian}\
  \bibnamefont {Lu}}, \bibinfo {author} {\bibfnamefont {Thibault}\ \bibnamefont
  {Marin}}, \bibinfo {author} {\bibfnamefont {Yue}\ \bibnamefont {Zhuo}},
  \bibinfo {author} {\bibfnamefont {Yi-Fan}\ \bibnamefont {Chen}}, \ and\
  \bibinfo {author} {\bibfnamefont {Chao}\ \bibnamefont {Ma}},\ }\href@noop {}
  {\enquote {\bibinfo {title} {Accelerating {MRI} reconstruction on {TPUs}},}\
  } (\bibinfo {year} {2020}{\natexlab{a}}),\ \Eprint
  {http://arxiv.org/abs/2006.14080} {arXiv:2006.14080 [cs.CE]} \BibitemShut
  {NoStop}%
\bibitem [{\citenamefont {Ma}\ \emph {et~al.}(2021)\citenamefont {Ma},
  \citenamefont {Marin}, \citenamefont {Lu}, \citenamefont {fan Chen},\ and\
  \citenamefont {Zhuo}}]{TPUFFT1}%
  \BibitemOpen
  \bibfield  {author} {\bibinfo {author} {\bibfnamefont {Chao}\ \bibnamefont
  {Ma}}, \bibinfo {author} {\bibfnamefont {Thibault}\ \bibnamefont {Marin}},
  \bibinfo {author} {\bibfnamefont {TJ}~\bibnamefont {Lu}}, \bibinfo {author}
  {\bibfnamefont {Yi}~\bibnamefont {fan Chen}}, \ and\ \bibinfo {author}
  {\bibfnamefont {Yue}\ \bibnamefont {Zhuo}},\ }\bibfield  {title} {\enquote
  {\bibinfo {title} {Nonuniform fast fourier transform on {TPUs}},}\ \
  }(\bibinfo {year} {2021})\BibitemShut {NoStop}%
\bibitem [{\citenamefont {Lu}\ \emph {et~al.}(2020{\natexlab{b}})\citenamefont
  {Lu}, \citenamefont {Chen}, \citenamefont {Hechtman}, \citenamefont {Wang},\
  and\ \citenamefont {Anderson}}]{TPUFFT2}%
  \BibitemOpen
  \bibfield  {author} {\bibinfo {author} {\bibfnamefont {Tianjian}\
  \bibnamefont {Lu}}, \bibinfo {author} {\bibfnamefont {Yi-Fan}\ \bibnamefont
  {Chen}}, \bibinfo {author} {\bibfnamefont {Blake}\ \bibnamefont {Hechtman}},
  \bibinfo {author} {\bibfnamefont {Tao}\ \bibnamefont {Wang}}, \ and\ \bibinfo
  {author} {\bibfnamefont {John}\ \bibnamefont {Anderson}},\ }\href@noop {}
  {\enquote {\bibinfo {title} {Large-scale discrete fourier transform on
  {TPUs}},}\ } (\bibinfo {year} {2020}{\natexlab{b}}),\ \Eprint
  {http://arxiv.org/abs/2002.03260} {arXiv:2002.03260 [cs.MS]} \BibitemShut
  {NoStop}%
\bibitem [{\citenamefont {Huot}\ \emph {et~al.}(2019)\citenamefont {Huot},
  \citenamefont {Chen}, \citenamefont {Clapp}, \citenamefont {Boneti},\ and\
  \citenamefont {Anderson}}]{huot2019highresolution}%
  \BibitemOpen
  \bibfield  {author} {\bibinfo {author} {\bibfnamefont {Fantine}\ \bibnamefont
  {Huot}}, \bibinfo {author} {\bibfnamefont {Yi-Fan}\ \bibnamefont {Chen}},
  \bibinfo {author} {\bibfnamefont {Robert}\ \bibnamefont {Clapp}}, \bibinfo
  {author} {\bibfnamefont {Carlos}\ \bibnamefont {Boneti}}, \ and\ \bibinfo
  {author} {\bibfnamefont {John}\ \bibnamefont {Anderson}},\ }\href@noop {}
  {\enquote {\bibinfo {title} {High-resolution imaging on {TPUs}},}\ }
  (\bibinfo {year} {2019}),\ \Eprint {http://arxiv.org/abs/1912.08063}
  {arXiv:1912.08063 [cs.CE]} \BibitemShut {NoStop}%
\bibitem [{\citenamefont {et~al.}()}]{tpu_algebra}%
  \BibitemOpen
  \bibfield  {author} {\bibinfo {author} {\bibfnamefont {Adam G. M.~Lewis}\
  \bibnamefont {et~al.}},\ }\href@noop {} {\enquote {\bibinfo {title} {{Tensor
  Processing Units for Distributed Dense Linear Algebra}},}\ }\bibinfo
  {howpublished} {{Sandbox@Alphabet, \textit{in preparation}.}}\BibitemShut
  {Stop}%
\bibitem [{\citenamefont {{Martin Ganahl et al.}}()}]{tpu_circuit}%
  \BibitemOpen
  \bibfield  {author} {\bibinfo {author} {\bibnamefont {{Martin Ganahl et
  al.}}},\ }\href@noop {} {\enquote {\bibinfo {title} {{Tensor Processing Units
  for Simulating Quantum Circuits}},}\ }\bibinfo {howpublished}
  {{Sandbox@Alphabet, \textit{in preparation}.}}\BibitemShut {Stop}%
\bibitem [{\citenamefont {Morningstar}\ \emph {et~al.}()\citenamefont
  {Morningstar}, \citenamefont {Hauru}, \citenamefont {Beall}, \citenamefont
  {Ganahl}, \citenamefont {Lewis}, \citenamefont {Khemani},\ and\ \citenamefont
  {Vidal}}]{tpu_floquet}%
  \BibitemOpen
  \bibfield  {author} {\bibinfo {author} {\bibfnamefont {Alan}\ \bibnamefont
  {Morningstar}}, \bibinfo {author} {\bibfnamefont {Markus}\ \bibnamefont
  {Hauru}}, \bibinfo {author} {\bibfnamefont {Jackson}\ \bibnamefont {Beall}},
  \bibinfo {author} {\bibfnamefont {Martin}\ \bibnamefont {Ganahl}}, \bibinfo
  {author} {\bibfnamefont {Adam G.~M.}\ \bibnamefont {Lewis}}, \bibinfo
  {author} {\bibfnamefont {Vedika}\ \bibnamefont {Khemani}}, \ and\ \bibinfo
  {author} {\bibfnamefont {Guifre}\ \bibnamefont {Vidal}},\ }\href@noop {}
  {\enquote {\bibinfo {title} {{Simulation of quantum many-body dynamics with
  Tensor Processing Units: Floquet prethermalization}},}\ }\bibinfo
  {howpublished} {{Sandbox@Alphabet, \textit{in preparation}.}}\BibitemShut
  {Stop}%
\bibitem [{\citenamefont {Shillito}\ \emph {et~al.}()\citenamefont {Shillito},
  \citenamefont {Petrescu}, \citenamefont {Cohen}, \citenamefont {Beall},
  \citenamefont {Hauru}, \citenamefont {Ganahl}, \citenamefont {Lewis},
  \citenamefont {Blais},\ and\ \citenamefont {Vidal}}]{tpu_qhardware}%
  \BibitemOpen
  \bibfield  {author} {\bibinfo {author} {\bibfnamefont {Ross}\ \bibnamefont
  {Shillito}}, \bibinfo {author} {\bibfnamefont {Alexandru}\ \bibnamefont
  {Petrescu}}, \bibinfo {author} {\bibfnamefont {Joachim}\ \bibnamefont
  {Cohen}}, \bibinfo {author} {\bibfnamefont {Jackson}\ \bibnamefont {Beall}},
  \bibinfo {author} {\bibfnamefont {Markus}\ \bibnamefont {Hauru}}, \bibinfo
  {author} {\bibfnamefont {Martin}\ \bibnamefont {Ganahl}}, \bibinfo {author}
  {\bibfnamefont {Adam G.~M.}\ \bibnamefont {Lewis}}, \bibinfo {author}
  {\bibfnamefont {Alexandre}\ \bibnamefont {Blais}}, \ and\ \bibinfo {author}
  {\bibfnamefont {Guifre}\ \bibnamefont {Vidal}},\ }\href@noop {} {\enquote
  {\bibinfo {title} {{Classical simulation of superconducting quantum hardware
  using Tensor Processing Units }},}\ }\bibinfo {howpublished}
  {{Sandbox@Alphabet, \textit{in preparation}.}}\BibitemShut {Stop}%
\bibitem [{\citenamefont {{Ryan Pederson et al.}}()}]{tpu_qchem}%
  \BibitemOpen
  \bibfield  {author} {\bibinfo {author} {\bibnamefont {{Ryan Pederson et
  al.}}},\ }\href@noop {} {\enquote {\bibinfo {title} {{Tensor Processing Units
  for Quantum Chemistry }},}\ }\bibinfo {howpublished} {{Sandbox@Alphabet,
  \textit{in preparation}.}}\BibitemShut {Stop}%
\bibitem [{\citenamefont {{Ruyi Song et al.}}()}]{tpu_spinLED}%
  \BibitemOpen
  \bibfield  {author} {\bibinfo {author} {\bibnamefont {{Ruyi Song et al.}}},\
  }\href@noop {} {\enquote {\bibinfo {title} {{Simulation of Spin Light
  Emitting Diodes using Tensor Processing Units}},}\ }\bibinfo {howpublished}
  {{Sandbox@Alphabet, \textit{in preparation}.}}\BibitemShut {Stop}%
\bibitem [{\citenamefont {{Martin~Ganahl et al.}}()}]{tpu_DMRG}%
  \BibitemOpen
  \bibfield  {author} {\bibinfo {author} {\bibnamefont {{Martin~Ganahl et
  al.}}},\ }\href@noop {} {\enquote {\bibinfo {title} {{Density Matrix
  Renormalization Group using Tensor Processing Units}},}\ }\bibinfo
  {howpublished} {{Sandbox@Alphabet, \textit{in preparation}.}}\BibitemShut
  {Stop}%
\bibitem [{\citenamefont {Gustafson}\ \emph {et~al.}(2021)\citenamefont
  {Gustafson}, \citenamefont {Holzman}, \citenamefont {Kowalkowski},
  \citenamefont {Lamm}, \citenamefont {Li}, \citenamefont {Perdue},
  \citenamefont {Boixo}, \citenamefont {Isakov}, \citenamefont {Martin},
  \citenamefont {Thomson} \emph {et~al.}}]{tpu_Z2field}%
  \BibitemOpen
  \bibfield  {author} {\bibinfo {author} {\bibfnamefont {Erik}\ \bibnamefont
  {Gustafson}}, \bibinfo {author} {\bibfnamefont {Burt}\ \bibnamefont
  {Holzman}}, \bibinfo {author} {\bibfnamefont {James}\ \bibnamefont
  {Kowalkowski}}, \bibinfo {author} {\bibfnamefont {Henry}\ \bibnamefont
  {Lamm}}, \bibinfo {author} {\bibfnamefont {Andy C.~Y.}\ \bibnamefont {Li}},
  \bibinfo {author} {\bibfnamefont {Gabriel}\ \bibnamefont {Perdue}}, \bibinfo
  {author} {\bibfnamefont {Sergio}\ \bibnamefont {Boixo}}, \bibinfo {author}
  {\bibfnamefont {Sergei}\ \bibnamefont {Isakov}}, \bibinfo {author}
  {\bibfnamefont {Orion}\ \bibnamefont {Martin}}, \bibinfo {author}
  {\bibfnamefont {Ross}\ \bibnamefont {Thomson}},  \emph {et~al.},\ }\href@noop
  {} {\enquote {\bibinfo {title} {Large scale multi-node simulations of
  $\mathbb{Z}_2$ gauge theory quantum circuits using google cloud platform},}\
  } (\bibinfo {year} {2021}),\ \Eprint {http://arxiv.org/abs/2110.07482}
  {arXiv:2110.07482 [quant-ph]} \BibitemShut {NoStop}%
\bibitem [{Not()}]{Note_Physics_1}%
  \BibitemOpen
  \href@noop {} {}\bibinfo {howpublished} {All the results presented in this
  paper correspond to end-to-end computations conducted in single precision.
  This is in contrast with the double precision often used in scientific
  computing. Our numerical experiments confirmed that single precision, capable
  of roughly 7 digits of accuracy, was sufficient to yield stable, numerically
  precise results. This is consistent with the error accumulation analysis of
  Ref.~\onlinecite{tpu_floquet}, where very deep quantum circuits, with several
  millions of gates, were successfully simulated using single
  precision.}\BibitemShut {Stop}%
\bibitem [{\citenamefont {Bradbury}\ \emph {et~al.}(2018)\citenamefont
  {Bradbury}, \citenamefont {Frostig}, \citenamefont {Hawkins}, \citenamefont
  {Johnson}, \citenamefont {Leary}, \citenamefont {Maclaurin}, \citenamefont
  {Necula}, \citenamefont {Paszke}, \citenamefont {Vander{P}las}, \citenamefont
  {Wanderman-{M}ilne},\ and\ \citenamefont {Zhang}}]{jax}%
  \BibitemOpen
  \bibfield  {author} {\bibinfo {author} {\bibfnamefont {James}\ \bibnamefont
  {Bradbury}}, \bibinfo {author} {\bibfnamefont {Roy}\ \bibnamefont {Frostig}},
  \bibinfo {author} {\bibfnamefont {Peter}\ \bibnamefont {Hawkins}}, \bibinfo
  {author} {\bibfnamefont {Matthew~James}\ \bibnamefont {Johnson}}, \bibinfo
  {author} {\bibfnamefont {Chris}\ \bibnamefont {Leary}}, \bibinfo {author}
  {\bibfnamefont {Dougal}\ \bibnamefont {Maclaurin}}, \bibinfo {author}
  {\bibfnamefont {George}\ \bibnamefont {Necula}}, \bibinfo {author}
  {\bibfnamefont {Adam}\ \bibnamefont {Paszke}}, \bibinfo {author}
  {\bibfnamefont {Jake}\ \bibnamefont {Vander{P}las}}, \bibinfo {author}
  {\bibfnamefont {Skye}\ \bibnamefont {Wanderman-{M}ilne}}, \ and\ \bibinfo
  {author} {\bibfnamefont {Qiao}\ \bibnamefont {Zhang}},\ }\href
  {http://github.com/google/jax} {\enquote {\bibinfo {title} {{JAX}: composable
  transformations of {P}ython+{N}um{P}y programs},}\ } (\bibinfo {year}
  {2018})\BibitemShut {NoStop}%
\bibitem [{XLA()}]{XLA}%
  \BibitemOpen
  \href@noop {} {}\bibinfo {howpublished} {\url{https://tensorflow.org/xla}},\
  \bibinfo {note} {accessed: 2021-10-01}\BibitemShut {NoStop}%
\bibitem [{\citenamefont {Lam}\ \emph {et~al.}(2015)\citenamefont {Lam},
  \citenamefont {Pitrou},\ and\ \citenamefont {Seibert}}]{numba}%
  \BibitemOpen
  \bibfield  {author} {\bibinfo {author} {\bibfnamefont {Siu~Kwan}\
  \bibnamefont {Lam}}, \bibinfo {author} {\bibfnamefont {Antoine}\ \bibnamefont
  {Pitrou}}, \ and\ \bibinfo {author} {\bibfnamefont {Stanley}\ \bibnamefont
  {Seibert}},\ }\bibfield  {title} {\enquote {\bibinfo {title} {Numba: A
  llvm-based python jit compiler},}\ }in\ \href {\doibase
  10.1145/2833157.2833162} {\emph {\bibinfo {booktitle} {Proceedings of the
  Second Workshop on the LLVM Compiler Infrastructure in HPC}}},\ \bibinfo
  {series and number} {LLVM '15}\ (\bibinfo  {publisher} {Association for
  Computing Machinery},\ \bibinfo {address} {New York, NY, USA},\ \bibinfo
  {year} {2015})\BibitemShut {NoStop}%
\bibitem [{\citenamefont {{Henkel}}(1999)}]{henkel_conformal_1999}%
  \BibitemOpen
  \bibfield  {author} {\bibinfo {author} {\bibfnamefont {Malte}\ \bibnamefont
  {{Henkel}}},\ }\href {http://link.springer.com/10.1007/978-3-662-03937-3}
  {\emph {\bibinfo {title} {Conformal {Invariance} and {Critical}
  {Phenomena}}}}\ (\bibinfo  {publisher} {{Springer}},\ \bibinfo {address}
  {Berlin, {Heidelberg}},\ \bibinfo {year} {1999})\BibitemShut {NoStop}%
\bibitem [{\citenamefont {Hayden}\ \emph {et~al.}(2006)\citenamefont {Hayden},
  \citenamefont {Leung},\ and\ \citenamefont {Winter}}]{Hayden2006}%
  \BibitemOpen
  \bibfield  {author} {\bibinfo {author} {\bibfnamefont {Patrick}\ \bibnamefont
  {Hayden}}, \bibinfo {author} {\bibfnamefont {Debbie}\ \bibnamefont {Leung}},
  \ and\ \bibinfo {author} {\bibfnamefont {Andreas}\ \bibnamefont {Winter}},\
  }\bibfield  {title} {\enquote {\bibinfo {title} {Aspects of generic
  entanglement},}\ }\href {\doibase 10.1007/s00220-006-1535-6} {\bibfield
  {journal} {\bibinfo  {journal} {Communications in Mathematical Physics}\
  }\textbf {\bibinfo {volume} {265}},\ \bibinfo {pages} {95--117} (\bibinfo
  {year} {2006})}\BibitemShut {NoStop}%
\bibitem [{\citenamefont {Kim}\ and\ \citenamefont {Huse}(2013)}]{KimHuse2013}%
  \BibitemOpen
  \bibfield  {author} {\bibinfo {author} {\bibfnamefont {Hyungwon}\
  \bibnamefont {Kim}}\ and\ \bibinfo {author} {\bibfnamefont {David~A.}\
  \bibnamefont {Huse}},\ }\bibfield  {title} {\enquote {\bibinfo {title}
  {Ballistic spreading of entanglement in a diffusive nonintegrable system},}\
  }\href {\doibase 10.1103/PhysRevLett.111.127205} {\bibfield  {journal}
  {\bibinfo  {journal} {Phys. Rev. Lett.}\ }\textbf {\bibinfo {volume} {111}},\
  \bibinfo {pages} {127205} (\bibinfo {year} {2013})},\ \Eprint
  {http://arxiv.org/abs/1306.4306} {arXiv:1306.4306} \BibitemShut {NoStop}%
\end{thebibliography}%

\appendix


\section{Distribution and manipulation of an $N$-qubit wavefunction on TPUs}
\label{app:distribution}

Consider the wavefunction $\ket{\Psi}$ of a system of $N$ qubits,
\begin{equation}
    \ket{\Psi} = \sum_{b_1=0}^1 \sum_{b_2=0}^1 \cdots \sum_{b_N=0}^1 \Psi_{b_1 b_2 \cdots b_N} \ket{b_1 b_2 \cdots b_N},
\end{equation}
which is characterized by $2^N$ complex amplitudes $\Psi_{b_1 b_2 \cdots b_N} = \langle b_1 b_2 \cdots b_N|\Psi \rangle$. In this appendix we explain how to distribute the $2^{N}$ amplitudes $\Psi_{b_1 b_2 \cdots b_N}$ over a number $2^{N_g}$ of TPU cores. We also explain how to update the distributed wavefunction under the action of a 7-qubit operator $h$, $\ket{\Psi} \mapsto h \ket{\Psi}$.

We index each amplitude $\Psi_{b_1 b_2 \cdots b_N}$ by a bit-string $(b_1, \dots, b_N)$, where $b_i \in \{0, 1\}$, corresponding to the computational basis vector $\ket{b_1 b_2 \cdots b_N}$. We then divide the $N$ qubits into two groups. The first group contains $N_g$ qubits, which we call \textit{global} qubits. The second group contains the remaining $N_l = N - N_g$ qubits, referred to as \textit{local} qubits. Accordingly, each bit-string $(b_1, \dots, b_N)$ is divided into a global part $(b_1, \dots, b_{N_g})$ and local part  $(b_{N_g + 1}, \dots, b_N)$. The global part  $(b_1, \dots, b_{N_g})$ determines the TPU core where the amplitude is stored. For instance, for $N_g=3$ we have 8 cores, labelled $\{(000),(001),(010), \cdots, (111) \}$, and core $(000)$ stores the amplitudes $\Psi_{000b_4\cdots b_N}$, core $(001)$ stores the amplitudes $\Psi_{001b_4\cdots b_N}$, etc. In this way, core $(c_1, \cdots, c_{N_g})$ (where we temporarily label the global part of the bit-string using $c$'s to emphasize that they are constant on a given core) stores the $2^{N_l}$-element sub-array $\Psi_{c_1\cdots c_{N_g}b_{N_g+1}\cdots b_{N}}$ with components labelled by the local bit-string $(b_{N_g+1},\cdots, b_{N})$ corresponding to the local qubits.

In other words, we think of the $2^N$ amplitudes $\Psi_{b_1b_2\cdots b_N}$ of a single distributed array as $2^{N_g}$ sub-arrays of size $2^{N_l}$, each one stored locally on the corresponding TPU core. Graphically we denote this with the following diagram, where the lines denote qubits with the red ones on the left being the global ones:
\begin{align*}
\includegraphics[width=0.9\linewidth]{wavefunction_local_global.pdf}
\end{align*}

Data in TPU memory is stored in chunks of size $8 \times 128$. This means that, for any array, the last two indices must have sizes that are multiples of $8$ and $128$ (otherwise the array we will be padded with zeros until it has that shape, incurring a waste of memory that should be avoided). Given that $2^3=8$ and $2^{7}=128$, this corresponds to using the last two indices of the array to label at least 3 and 7 qubits, respectively. We thus store the wavefunction using a distributed array with a shape that adjusts to this rule. For instance, with the shape ($2^{N_g}, 2^{N_l-10}, 2^3,2^7$), that is, where each local sub-array on each core has shape ($ 2^{N_l-10}, 2^3,2^7$). Graphically:
\begin{align*}
\includegraphics[width=0.9\linewidth]{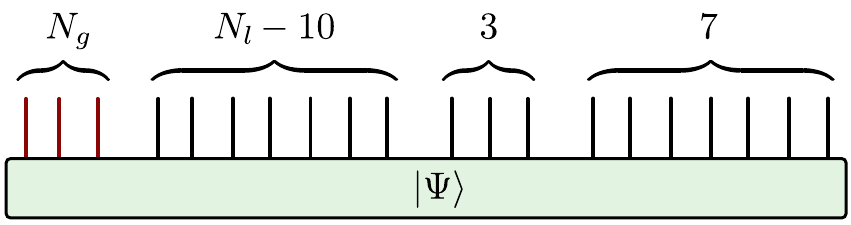}
\end{align*}

Our next step is to consider the update $\ket{\Psi} \mapsto h\ket{\Psi}$, where $h$ represents a 7-qubit operator, or a matrix with shape $(128,128)$, which we broadcast so that each core holds a copy. Let us label the components $h_{\vec{\alpha} \vec{\alpha}'}$ of this matrix by indices $\vec{\alpha}$ resulting from composing 7 binary indices, $\vec{\alpha}=(\alpha_1,\alpha_2, \cdots, \alpha_7)$, with $\alpha_i = \{0,1\}$. We next consider three cases, depending on which qubits the 7-qubit operator $h$ acts on. 

(i) First we assume that $h$ acts on the last 7 local qubits. We can then simply multiply $h$ and the local $\ket{\Psi}$ together by contracting over the last index of the local sub-array,
\begin{eqnarray}
    &&\Psi_{c_1\cdots c_{N_g}b_{N_g+1} \cdots b_{N-7} \vec{\alpha}} \mapsto  \\
    \sum_{\vec{\alpha}'} h_{\vec{\alpha}, \vec{\alpha}'} ~&&\Psi_{c_1\cdots c_{N_g}b_{N_g+1} \cdots b_{N-7} \vec{\alpha}'}
\end{eqnarray}
Graphically this can be represented as

\begin{align*}
\includegraphics[width=0.9\linewidth]{wavefunction_op_product_last_qubits.pdf}
\end{align*}

Importantly, this update of the distributed wavefunction can be accomplished by having each TPU core update its corresponding local sub-array, so that the $2^{N_g}$ local sub-arrays are updated in parallel without need for inter-core communication. 

(ii) Next we assume that the 7-qubit gate acts on arbitrary local qubits, and not just the last 7 qubits as before. In this case, we re-organize each of the local sub-arrays so as to move the 7 targeted qubits to the the last 7 positions, followed by the above contraction, and then by the reversal of the above re-organization. The reorganization of a local sub-array consists of a sequence of reshapes and transpositions. For instance, we could reshape the local sub-array from shape ($2^6, 2^3, 2^7$) to shape ($2^2, 2^7, 2^7$), represented by
\begin{align*}
\includegraphics[width=0.9\linewidth]{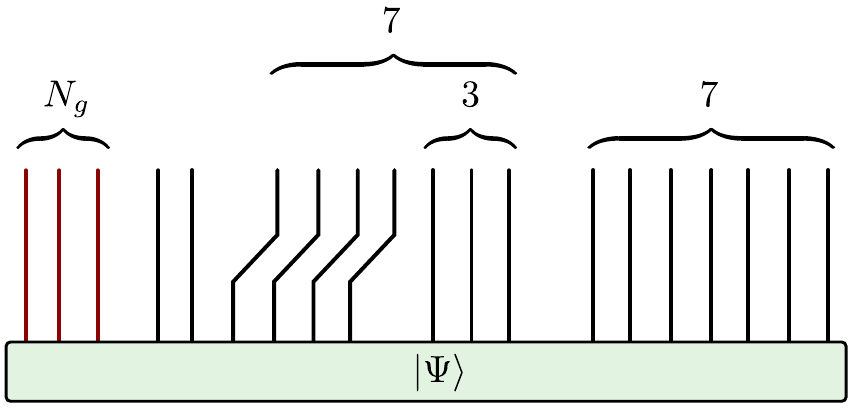}
\end{align*}
and then transpose the two last indices of the resulting local sub-array, which we draw as
\begin{align*}
\includegraphics[width=0.9\linewidth]{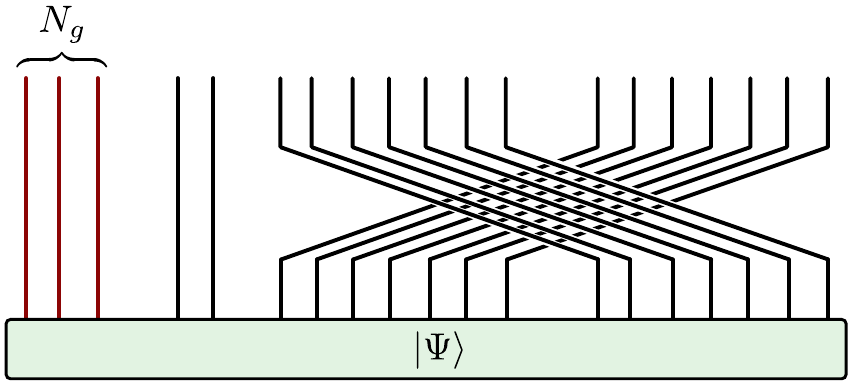}
\end{align*}
Notice that at each step of this example, we store a local array where the range of the last two indices correspond to, at least, 3 and 7 qubits, respectively (that is, to multiples of 8 and 128), so as to prevent padding with zeros. Once again, all these operations are performed in parallel without need for inter-core communication. 

(iii) Finally, to apply the update $\ket{\Psi} \mapsto h\ket{\Psi}$ when $h$ acts on a set of qubits that includes at least one of the global qubits, we first re-distribute the wavefunction so that all the targeted qubits become local, then proceed as above, and then re-distribute the resulting wavefunction back to the original form. For instance, if $N_g=3$ and $h$ acts on the three global qubits, but not on the first three local qubits, we can re-distribute the wavefunction by swapping the global qubits with the first three local qubits, represented as
\begin{align*}
\includegraphics[width=0.9\linewidth]{wavefunction_global_swap.pdf}
\end{align*}
Notice that this transformation cannot be performed locally on each TPU core, but requires instead substantial communication between the cores. TPUs can perform such inter-core communication remarkably fast, thanks to the dedicated ICIs that connects the cores directly, bypassing the CPU hosts on each TPU board.


\section{Local Hamiltonian}
\label{app:HPsi}

As discussed in the main text, TPUs natively do matrix products of size $128 \times 128$, or $2^7 \times 2^7$. In this Appendix we explain how to use this ability to compute the update $\ket{\Psi} \mapsto H \ket{\Psi}$ for a local Hamiltonian $H$ that decomposes as a sum of local terms.

Consider a local Hamiltonian
\begin{equation} \label{eq:local_Hamiltonian}
    H = \sum_{i=1}^Q k_{i}     
\end{equation}
where each of the $Q$ terms $k_{i}$ is an operator that acts non-trivially on at most 7 qubits, possibly less. Recall that on a TPU, a matrix multiplication involving matrices with dimensions less than $128 \times 128$ is done by padding the matrices with zeros until they are of size $128 \times 128$. That means that we would like to be working with local terms $h_i$ that act on exactly 7 qubits, not less. Accordingly, our first goal is to rewrite the above Hamiltonian as
\begin{equation}
    H = \sum_{i=1}^P h_{i}     
\end{equation}
where each 7-qubit term $h_i$ may come from adding two or more terms $k_i$ in Eq. \eqref{eq:local_Hamiltonian}. As an example, consider a 1D local Hamiltonian where each term $k_i$ acts on two consecutive qubits, denoted $k_{[i,i+1]}$. Then we define the 7-qubit terms $h_i$, denoted $h_{[i,i+6]}$, by adding together 6 of such terms. For instance, the first 7-qubit operator reads
\begin{align}
\label{eq:first_k}
\begin{split}
     h_{[1,7]} =& \;k_{[1,2]} \otimes \unity_5 + \unity_1 \otimes k_{[2,3]} \otimes \unity_4 +\\%
    &\; \unity_2 \otimes k_{[3,4]} \otimes \unity_3 +\dots + \unity_5 \otimes k_{[6,7]}.
\end{split}
\end{align}
Here $\unity_m$ is the identity on $m$ qubits. 
Graphically we represent this as
\begin{align*}
\includegraphics[width=0.9\linewidth]{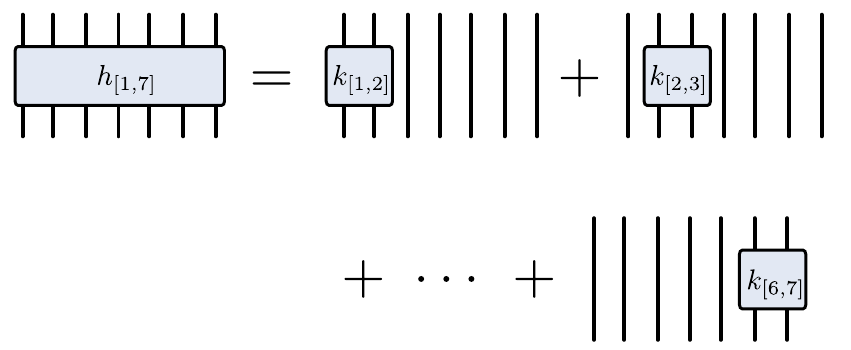}
\end{align*}
We can similarly define $h_{[7,13]}$, $h_{[13,19]}$, etc. Notice that, in this example, if e.g. $N=22$ , the last term $h_{[19,22]}$ only acts on 4 qubits. In that case, we will tensor product $h_{[19,22]}$ with the identity $\unity_3$ on 3 additional qubits, into a 7-qubit operator $h_{[16,22]} = \unity_3 \otimes h_{[19,22]}$.

To compute $\ket{\Psi} \mapsto H\ket{\Psi}$ for $H = \sum_i h_i$, we compute a sequence of products $h_i \ket{\Psi}$ for $i=1,\cdots,P$, which we accumulate in wavefunctions $\ket{\Psi^{(i)}}$,
\begin{equation}
    \ket{\Psi^{(i)}} = \ket{\Psi^{(i-1)}} + h_{i} \ket{\Psi},~~~~i=1,\cdots, P, 
\end{equation}
where $\ket{\Psi^{(0)}} =0 $ is the null vector and $\ket{\Psi^{(P)}} = H \ket{\Psi}$ contains the final result. Notice that during the above steps we need to keep at least three distributed arrays on the TPUs, corresponding to wavefunctions $\Psi$, $h_i\ket{\Psi}$ and $\ket{\Psi^{(i-1)}}$.

It is instructive to evaluate what the loss in efficiency is if we disregard the rule that matrix products should involve a matrices of shape $(2^7,2^7) = (128, 128)$ (or, more generally, multiples of these dimensions). Suppose, as a concrete example, that our goal is to multiply the $2^{N}$-element distributed array for an N-qubit wavefunction $\ket{\Psi}$ [properly reshaped into a matrix with shape $(2^{N-2}, 2^{2})$] by a two-qubit term $k$ [or matrix with shape $(2^2,2^2)$]. On a CPU, this matrix product would require roughly $2^{N-2}\times 2^2 \times 2^2 = 2^{N+2}$ floating-point multiplications (and a similar number of additions, which we ignore in our counting for simplicity). On a TPU, we first need to pad both matrices with zeros until they reach shapes $(2^{N-2},2^{7})$ and $(2^{7},2^7)$, respectively. We see that storing the wavefunction in this format requires the same memory as $2^5=32$ copies of $\ket{\Psi}$ stored without padding. For large $N$, this is a massive waste of memory! In addition, the matrix multiplication now requires $2^{N-2}\times 2^7 \times 2^7 = 2^{N+12}$ floating-point multiplications, which is $2^{10}=1024 \times$ more than on a CPU! Again, a massive waste of compute time.

Given the constraints forced on us by the MXU, how can we more efficiently accomplish the above product? We proceed by augmenting the two-qubit operator $k$ into a 7-qubit operator $h$, given by $h = k \otimes \unity_5$, that acts as the identity on 5 additional qubits. We then reshape the $2^N$-element array into a matrix of shape $(2^{N-7},2^7)$ and multiply it by the $(2^7,2^7)$-shaped matrix for $h$. Notice that no padding with zeros is now required. In addition, the matrix multiplication now requires roughly $2^{N-7}\times 2^7 \times 2^7 = 2^{N+7}$ floating-point operations, which is $2^5=32\times$ more the operations needed on a CPU (still a very significant waste, but better than the factor 1024$\times$). In addition, If we can join the two-qubit term $k$ with other two-qubit terms that are also part of the same Hamiltonian, we can include them in the same matrix multiplication. For instance, in Eq. \eqref{eq:first_k} we joined 6 two-qubit terms into a single seven-qubit term. In that case the MXU forces us into a factor $32/6 \approx 5.3 \times$ more operations compared to a CPU. Needless to say, the extreme efficiency of the MXU by far compensates for the generation of additional floating-point operations. Finally, we remark that for 7-qubit operators (or operators acting on even a larger number of qubits), the MXU does not force us into generating additional floating-point operations, achieving its optimal efficiency.


\section{Expansion of the time evolution operator}
\label{app:time-evolution-expansion}
To numerically time evolve a quantum state, we take the time evolution operator for a small time $\delta t$, $\exp(-i \, \delta t \, H)$, and Taylor expand the exponential to sixth order:
\begin{align}
\exp x = \sum_{n=0}^6 \frac{1}{n!}x^n + O(x^7) = \prod_{n=1}^6 (1 + a_n x) + O(x^7)
\end{align}
The latter form is nothing but a rewriting of the polynomial $\sum_{n=0}^6 \frac{1}{n!}x^n$ in terms of its roots $a_n$, which can be solved for numerically.
To accuracy sufficient for single precision, the roots are

\begin{align*}
a_1 &= 0.37602583 - 0.13347447i,~~~~~~ a_2 = a_1^*,\\%
a_3 &= -0.05612287 - 0.25824122i,~~~~ a_4 = a_3^*,\\%
a_5 &= 0.18009704 + 0.30409897i,~~~~~~ a_6 = a_5^*,\\%
\end{align*}

\section{Memory efficient Lanczos algorithm}
\label{app:lanczos}
The Lanczos method is a matrix-free diagonalization technique which can be used to approximate extremal (largest in magnitude) eigenvector-eigenvalue pairs of a large Hermitian linear operator $H$. The only requirements (besides $H$ being Hermitian) is the ability to efficiently compute the action of $H$ on a given vector $\x$, and enough memory to store at least four such vectors (plus some small amount of extra memory to perform auxiliary computations).

\begin{figure}
\begin{algorithm}[H]
    \caption{Lanczos algorithm}\label{alg:lanczos}
    \begin{algorithmic}[1]
    \Function{Lanczos($H, \x, K, \delta$)}{}
    \State $\x_{-1} = \rm{zeros\_like}(\x)$
    \State $\x_0 = \x$
    \State $Q_{-1}$ = []
    \For{$n$=0\dots $K$-1}
        \State $\beta_n = \lVert \x_n \rVert$
        \If {$\beta_n<\delta$} \Comment{invariant subspace found}
            \State \Return $\{\alpha_0,\dots,\alpha_{n-1}\},\{\beta_0,\dots,\beta_{n}\}, Q_{n-1}$
        \EndIf
        \State $\x_n \gets \x_n/\beta_n$
        \State $Q_n = [Q_{n-1}, \x_n]$
        \State $\x_{n+1} = A\x_n$
        \State $\alpha_n = \langle \x_{n+1},\x_n\rangle$ \Comment{$\alpha_n\in \mathbbm{R} $}
        \State $\x_{n+1} \gets \x_{n+1} - \alpha_n \x_n - \beta_n \x_{n-1}$
    \EndFor
    \State \Return $\{\alpha_0,\dots,\alpha_{K-1}\},\{\beta_0,\dots,\beta_{K}\}, Q_{n-1}$
    \EndFunction
    \end{algorithmic}
\end{algorithm}
\end{figure}

\begin{figure}
\begin{algorithm}[H]
    \caption{Lanczos algorithm for computing a tri-diagonalization of $H$}\label{alg:lanczos_tridiag}
    \begin{algorithmic}[1]
    \Function{Lanczos\_tridiag($H, \x, K, \delta$)}{}
    \State $\x_{-1} = \rm{zeros\_like}(\x)$
    \State $\x_0 = \x$
    \For{$n$=0\dots $K$-1}
        \State $\beta_n = \lVert \x_0 \rVert$
        \If {$\beta_n<\delta$} \Comment{invariant subspace found}
            \State \Return $\{\alpha_0,\dots,\alpha_{n-1}\},\{\beta_0,\dots,\beta_{n}\}$
        \EndIf
        \State $\x_0 \gets \x_0/\beta_n$
        \State $\x_1 = A\x_0$
        \State $\alpha_n = \langle \x_1,\x_0\rangle$ \Comment{$\alpha_n\in \mathbbm{R} $}
        \State $\x_1 \gets \x_1 - \alpha_n \x_0 - \beta_n \x_{-1}$
        \State $\x_{-1} \gets \x_0$
        \State $\x_0\gets \x_1$
    \EndFor
    \State \Return $\{\alpha_0,\dots,\alpha_{K-1}\},\{\beta_0,\dots,\beta_{K}\}$
    \EndFunction
    \end{algorithmic}
\end{algorithm}
\end{figure}

\begin{figure}
\begin{algorithm}[H]
    \caption{Lanczos algorithm for computing an approximate ground state of $H$}\label{alg:lanczos_gs}
    \begin{algorithmic}[1]
    \Function{Lanczos\_groundstate($H, \x, \{\alpha_i\}, \{\beta_i\},\{v_i\}$)}{}
    \State $\mathbf{u} = \rm{zeros\_like}(\x)$
    \State $\x_{-1} = \rm{zeros\_like}(\x)$
    \State $\x_0 = \x$
    \For{$n$=0\dots\rm{len}($\{v_i\}$)} 
        \State $\x_0 \gets \x_0/\beta_n$
        \State $\mathbf{u} \gets \mathbf{u} + v_n \x_0$
        \State $\x_1 = A\x_0$
        \State $\x_1 \gets \x_1 - \alpha_n \x_0 - \beta_n \x_{-1}$
        \State $\x_{-1} \gets \x_0$
        \State $\x_0\gets \x_1$
    \EndFor
    \State \Return $\mathbf{u}$
    \EndFunction
    \end{algorithmic}
\end{algorithm}
\end{figure}

Algorithm \ref{alg:lanczos} outlines the basic Lanczos method. Starting from a normalized initial guess $\x$ for an extremal eigenstate of $H$, the Lanczos method iteratively builds an orthogonal basis $Q_{j} \equiv [\x_0, \x_1, \dots , \x_{j}]$ of Krylov vectors $\x_n$ for the Krylov-subspace $\rm{span}\{\x, H\x, \dots, H^{j}\x\}$. The algorithm makes explicit use of the Hermiticity of $H$ to obtain the orthogonal basis $Q_j$ through a three-step recurrence relation which at any point involves only three consecutive Krylov vectors. The Lanczos method converges towards the {\it dominant} eigenstate, i.e. the one with largest-in-magnitude eigenvalue, or a random linear superposition of states if the eigenvalue is degenerate. To achieve convergence towards the (algebraic) lowest eigenstate, one can e.g. apply a uniform spectral shift to $H$.
The algorithm produces the real coefficients $\{\alpha_n\}, \{\beta_n\}$ of the tridiagonal matrix $T_j$,
\begin{align}
T_j\equiv
    \left(
        \begin{array}{ccccc}
            \alpha_0&\beta_1&&\\
            \beta_1&\alpha_1&\beta_2&&\\
            &\beta_2&\alpha_2&\ddots&\\
            &&\ddots&\ddots&\beta_{j}\\
            &&&\beta_{j}&\alpha_{j}
        \end{array}
    \right)
\end{align}
with $j$ at most equal to $K-1$, where $K$ is the Krylov space dimension chosen by the user, but smaller if the algorithm terminates early (see below). The matrices $Q_{j}, T_{j}$ and $H$ satisfy the Lanczos relation 
\begin{equation}
    HQ_{j} = Q_{j}T_{j} + \beta_{j+1} \x_{j+1} \mathbf{e}_{j+1}^T
\end{equation}
with $\mathbf{e}_{j+1}$ the $j$+1-st euclidean basis-vector of dimension $j+1$.
The algorithm terminates early if an invariant subspace is found, i.e. if a newly generated Krylov vector $\x_{n}$ is a linear superposition of previous Krylov vectors. In this case the eigenvalues of $T_j$ and the corresponding Ritz vectors (see below) are exact eigenvalues and eigenvectors of $H$. Given an eigenvector $\mathbf{v}$ of $T_j$ to eigenvalue $\lambda$, the corresponding Ritz vector $\mathbf{u}$ is obtained by expanding $\mathbf{v}$ in the Krylov basis $Q_j$:
\begin{align}
    \mathbf{u} = Q_{j} \mathbf{v}.
\end{align}
The pair ($\lambda,\mathbf{u}$) is called a Ritz pair.
If the method does not terminate early, the Ritz pair ($\lambda, \mathbf{u}$) corresponding to an extremal eigenvector-eigenvalue pair ($\lambda, \mathbf{v}$) of $T_{j}$ is an approximate eigenvalue-eigenvector pair of $H$. A rough measure of the quality of this approximation is given by the residual norm
\begin{align}
    \lVert H\mathbf{u} - \lambda \mathbf{u}\rVert =& 
    \lVert HQ_{j} \mathbf{v} - Q_{j} T_{j} \mathbf{v}\rVert=\nonumber\\
    =&\beta_K |v_{j}|\nonumber,
\end{align}
which is given by the modulus of the last element in $\mathbf{v}$ times $\beta_K$. The quality of the approximation usually increases with Krylov dimension $K$, with typical values of $K$ ranging from a few dozens to a few hundreds.

The Lanczos method is infamously ill-conditioned, resulting in loss of orthogonality, in finite precision arithmetic, of the constructed Krylov basis $Q_{j}$ for modest sizes of $K$, and the appearance of spurious degeneracies (``ghost modes") in the spectrum of $T_{j}$. We note that for the case of only computing a single extremal eigenvector, ghost modes are of no particular concern.

Algorithm \ref{alg:lanczos} requires the storage of the constructed Krylov basis $Q_{j}$ in memory. For large values of $j$ this can become prohibitive. One can work around this issue by splitting Algorithm \ref{alg:lanczos} into two separate runs. During the first run one only computes the tridiagonal matrix $T_{j}$ {\it without} storing the Krylov vectors $Q_{j}$. This requires only enough memory to store three consecutive Krylov vectors $\x_n$ in memory. One then diagonalizes $T_{j}$ to obtain the extremal eigenpair $(\lambda, \mathbf{v})$. Finally, one uses the expansion coefficients $\mathbf{v}$ as input to a second Lanczos run during which the desired extremal Ritz vector $\mathbf{u}$ is accumulatively computed from the reconstructed Krylov basis. The second run requires enough memory to store four Krylov vectors. The two algorithms are outlined in Algorithms \ref{alg:lanczos_tridiag} and \ref{alg:lanczos_gs}. This approach of course comes at the expense of additional computational time required for the second run.
This latter, memory-efficient implementation of Lanczos is what we use to find the ground state of a 1D spin chain and obtain for instance the results in Fig.~\ref{fig:groundstate_both}.

\end{document}